# Leveraging AI for Direct Bystander Intervention Against Cyberbullying

PEINUAN QIN, School of Computing, National University of Singapore, Singapore
JITING CHENG, School of Computing, National University of Singapore, Singapore
JUNGUP LEE, Department of Social Work, National University of Singapore, Singapore
JUNTI ZHANG, Institute of Data Science, National University of Singapore, Singapore
ZHIXING LIU, School of Computing, National University of Singapore, Singapore
YI-CHIEH LEE, National University of Singapore, Singapore

Cyberbullying is a pervasive problem in online environments, causing substantial psychological harm to victims. Although bystander intervention has proven effective in mitigating its impact, motivating bystanders to engage in direct intervention remains a persistent challenge. Studies have suggested that difficulties in intervention skills and defending self-efficacy hinder bystanders from initiating direct intervention. To address this challenge, we introduced EmojiGen, an AI intervention tool designed to empower bystanders for direct intervention. EmojiGen enabled users to simply select an emoji as an intention clue, which subsequently combined the cyberbullying context to generate responses. In a between-subjects experiment involving 90 participants on a custom-built social media platform, we found that EmojiGen significantly increased the frequency of direct bystander interventions, both in supporting victims and in confronting perpetrators, driven by different factors. EmojiGen also increased the sense of knowing how to help and defending self-efficacy, while reducing perceived workload and anxiety associated with initiating intervention. The study contributed to the CSCW community through offering an effective direct bystander intervention method and providing design implications for future cyberbullying interventions.

## 1 Introduction

Cyberbullying is a widespread issue related to social media use [124], such as harassment [12, 122] and trolling [25], which inflicts considerable psychological distress on victims [71, 105]. To address it, cyberbullying mitigation research has garnered significant attention. Among these studies, cyberbullying bystanders, a group that widely witnesses online violence but rarely intervenes [9, 63, 88, 97], have been increasingly studied for their potential in preventing cyberbullying [31]. Typically, some prosocial intervention [19] designs were introduced to nudge bystanders toward more supportive actions [37, 116]. For example, Taylor et al. [116] and DiFranzo et al. [37] separately used empathy and responsibility design, leading to more supportive interventions from bystanders.

However, these designs have been shown to be effective for promoting *indirect intervention*, such as reporting or flagging content [83], while having limited facilitation effects in *direct intervention* [31, 116], such as confronting perpetrators or supporting victims [83]. As suggested by [24, 100, 118], one key factor affecting direct intervention is defending self-efficacy, the belief in one's ability to successfully intervene and help someone being bullied [24]. Intervention skills are also considered crucial, without which bystanders would not intervene even if other prerequisites [31], such as noticing interventions and taking the responsibility [24], are met. Some studies applied education programs [53, 72, 96] to promote cyberbullying intervention; nevertheless, they are criticized for the increasing self-efficacy and willingness to intervene, without actually promoting defensive behavior in actual events [20, 64, 87]. This indicates the need to explore effective measures to promote bystanders' direct intervention.

In response to these challenges, our focus has shifted to the rapidly growing domain of AI, which has already found extensive application in mitigating cyberbullying through online detection [70, 113], monitoring [4, 23], and enforcing social norms [22, 58, 108, 129]. These existing AI studies emphasized maintaining the order of the social network through automatic approaches rather than focusing on empowering bystanders. On the contrary, we value the potential of AI in promoting bystanders' intervention ability and defending self-efficacy. For instance, in mental health, Sharma et al. [109] utilized AI to facilitate the development of peer-to-peer mental health support, thereby significantly enhancing users' ability and self-efficacy to provide peer support. This provides a feasible path using AI to bridge the gap in bystanders' intervention skills and enhance their defending self-efficacy for direct intervention, although no empirical studies have yet proven its effectiveness.

To address this gap, we proposed EmojiGen, an AI-assisted intervention tool, aimed at leveraging AI to support bystanders' intervention skills and defending self-efficacy, and explore its effectiveness in promoting direct intervention. When encountering cyberbullying, rather than developing an entire response, the bystanders can simply select emojis, such as 😠, as effective indicators of intention [29, 45, 57, 81]. Afterward, the EmojiGen combines the cyberbullying context and generates responses. This can be used for both comforting victims and confronting perpetrators. To evaluate its effectiveness, we explored the following three research questions.

**RQ1:** How does EmojiGen affect bystander direct intervention frequency, including both supporting victims and resisting perpetrators?

**RQ2:** How does EmojiGen affect bystander perceptions (defending self-efficacy, sense of knowing how to help, personal responsibility, anxiety, and workload) in cyberbullying intervention?

**RQ3:** How do changes in bystanders' perceptions (e.g., defending self-efficacy) mediate the effect of EmojiGen on their direct intervention behavior in cyberbullying situations?

We conducted a between-subjects experiment with 90 participants on a custom-built social media platform featuring a collection of carefully crafted and validated posts by domain experts. The participants were assigned two conditions. Participants in $C_{EG}$ can use EmojiGen support for response construction, while those in $C_{NE}$ cannot. The sequence of posts and conditions was counterbalanced to mitigate order effects. Our findings indicate that EmojiGen significantly increased the frequency of direct intervention by bystanders, including comforting victims and resisting perpetrators. Meanwhile, EmojiGen enhanced bystanders' defending self-efficacy and their sense of knowing how to help, while reducing their perceived workload and anxiety associated with direct intervention. Further analysis revealed that EmojiGen impacts the direct intervention frequency through a path of enhancing their intervention skills and defending self-efficacy. Nevertheless, qualitative research revealed its limitations, emphasizing personalization and the expression of the user's authentic voice.

This research offers two primary contributions to the CSCW community:

(1) We proposed EmojiGen, which successfully addresses the research gap of promoting bystanders' direct intervention. Through using AI to support intervention skills and defending self-efficacy, this study enriches the implementation and understanding of previous cyberbullying intervention methods [58, 70, 107].

(2) We offered actionable design implications that outline priorities for future work, emphasizing the importance of enabling personalized and authentic responses, thereby extending ongoing CSCW research on supportive AI tools for collective online well-being.

## 2 Related Work

### 2.1 Cyberbullying and Bystanders

Cyberbullying is a form of repeated and deliberate online attacks against victims who often struggle to defend themselves [52]. Cyberbullying exhibits complex social dynamics [121] and includes rich media content such as text, photos, and videos [18, 79, 120] as part of the attack. Nowadays, cyberbullying has become a pervasive and significant concern. Reports indicate that 69% of adult social media users have witnessed mean or cruel behaviour towards others on social networking sites, while 19% of teenagers admit to engaging in harassment behaviour in the past year [77]. Cyberbullying poses significant risks, including psychological trauma, exacerbation of psychological burdens, and even damage to mental and physical well-being [71, 74, 75, 105].

As proposed by survey data [98], bystanders make up a large group and the prevalence (75%) of cyber-bystanders is substantially higher than that of cyber-bullies (11.2%) and cyber-victims (8.2%), with similar evidence from other studies [9, 88, 97, 98], suggesting a foundation for bystanders to mitigate cyberbullying. Additionally, bystanders play an important role in inhibiting bullying behaviors [35, 82], as they can provide a timely and effective end to bullying behavior and can weaken or alleviate the harm caused to the victim [99, 111]. Polanin et al. indicated that when bystanders intervene on behalf of the victim, they successfully abate victimization more than 50% of the time [99]. Despite this, only a minority of bystanders engage in intervention behaviors [38, 93], which is called the bystander effect [73]. Holfeld highlighted a phenomenon of responsibility distribution in which bystanders often believe that they do not need to help the victim because someone else will [56].

Given the large population of bystanders and their demonstrated effectiveness in stopping cyberbullying, existing studies in the CSCW community are dedicated to transforming this group into active agents in cyberbullying prevention.

### 2.2 Bystander Intervention in Cyberbullying

In this section, we start from a theoretical framework that explains how and why bystanders decide to intervene (Section 2.2.1). We then reviewed two main ways bystanders intervene in cyberbullying: indirect and direct intervention (Section 2.2.2). Subsequently, we introduced some common prosocial designs and their effectiveness in promoting indirect intervention, but also their limitations in promoting direct intervention (Section 2.2.3). Next, we emphasized how intervention skills and defending self-efficacy serve as core design constructs to promote direct intervention (Section 2.2.4). Finally, we presented existing research that employs educational and training approaches to strengthen intervention skills and defensive self-efficacy, and we also discussed the limitations of these approaches. (Section 2.2.5).

*2.2.1 Bystander Intervention Model.* To theoretically ground bystander intervention, prior work has proposed the Bystander Intervention Model (BIM) [31]. The BIM conceptualizes intervention as a sequential decision-making process requiring bystanders to (1) notice the event, (2) interpret it as an emergency situation, (3) accept personal responsibility, (4) possess the necessary knowledge and skills to intervene, and (5) ultimately take action. Failure to intervene may occur at any stage of this process, even when bystanders hold prosocial intentions.

Beyond its roots in social psychology, BIM has also informed CSCW research. In cyberbullying contexts, prior studies have operationalized different BIM stages through interface and platform design [37, 116] to mitigate cyberbullying. These studies suggest BIM is not only a model of individual decision-making, but can guide the design of cyberbullying interventions. However, they have primarily focused on earlier or motivational BIM stages, such as fostering empathy or increasing perceived responsibility, rather than the later capability-oriented stage of whether bystanders have the necessary knowledge and skills to intervene. This helps explain why existing designs have been more successful in promoting indirect intervention than direct intervention, which typically requires bystanders to formulate contextually appropriate and socially risky responses in the moment.

*2.2.2   Indirect and Direct Bystander Intervention.* Prior research commonly distinguishes between two forms of bystander intervention in cyberbullying contexts: indirect and direct interventions [40, 56, 83]. *Indirect intervention* typically includes actions such as flagging, disliking, or reporting abusive content to platform authorities. These behaviors can reduce the visibility of harmful content and contribute to norm reinforcement without requiring direct engagement with the bully or victim [5, 11, 37]. In contrast, *direct intervention* involves bystanders actively engaging in the situation, such as comforting the victim, publicly defending them, or directly confronting the bully [3, 36, 99]. Direct intervention has been shown to be particularly effective: it can immediately disrupt bullying behavior, provide emotional support to victims, and signal clear social disapproval to both perpetrators and other bystanders [5, 33, 99]. Moreover, visible defending behaviors may inspire additional bystanders to intervene, amplifying their protective effect. Despite its effectiveness, direct intervention occurs far less frequently than indirect forms [38, 93].

*2.2.3   Prosocial Design and the Limits of Promoting Direct Intervention.* To encourage bystander intervention, many studies adopt prosocial design approaches, which aim to motivate users to act in socially positive ways, such as supporting victims or discouraging harmful behavior [37, 106, 116]. These designs often target indirect intervention behaviors by shaping social norms or increasing perceived responsibility. For example, DiFranzo et al. [37] increased bystanders' reporting behaviors by making audience size and social transparency salient, thereby fostering a sense of public accountability. Similarly, Taylor et al. [116] demonstrated that embedding empathy cues into platform design could increase bystanders' willingness to intervene. While effective in promoting indirect actions, these approaches largely avoid requiring bystanders to directly engage with victims or bullies. However, such prosocial designs have shown limited success in promoting direct intervention. Direct intervention is consistently perceived as more difficult and risky, as it requires bystanders to publicly articulate support or confrontation in emotionally charged and socially uncertain situations [38, 67]. Importantly, prior research suggests that this difficulty is not merely motivational but is closely tied to bystanders' perceived intervention competence.

*2.2.4   The Central Role of Intervention Skills and Defending Self-Efficacy.* Specifically, direct intervention places high demands on both intervention skills and defending self-efficacy [17, 24, 30, 100]. Intervention skills [17, 30] refer to bystanders' ability to formulate appropriate responses in cyberbullying situations, such as composing supportive and empathetic messages to victims or assertive yet non-escalatory confrontations toward bullies. These skills require nuanced judgment of tone, content, and timing, often under conditions of emotional stress and social uncertainty. Prior studies show that untrained individuals frequently struggle to produce effective supportive responses, even when they intend to help, resulting in hesitation or complete inaction [17, 30, 68, 85].

Closely intertwined with intervention skills is defending self-efficacy [24, 100], defined as bystanders' belief in their capability to successfully intervene on behalf of a victim [24]. Rooted in social cognitive theory, self-efficacy reflects individuals' perceived competence in performing specific actions under challenging circumstances [7]. In cyberbullying contexts, low defending self-efficacy has been shown to predict avoidance behaviors, as bystanders anticipate stress, social backlash, or failure when intervening publicly [47, 51]. Empirical studies consistently demonstrate that defending self-efficacy is a strong predictor of actual defending behaviors [24, 100, 117].

Taken together, prior work suggests that the persistent rarity of direct bystander intervention is not primarily a problem of prosocial motivation, but rather a problem of capability. Specifically, insufficient intervention skills and low defending self-efficacy might jointly constrain bystanders' ability to act, even when they are willing to help. Under such circumstances, the importance of approaches that can meaningfully strengthen both factors rather than merely encouraging prosocial intent.

*2.2.5 Education and Training for Enhancing Intervention Skills and Defending Self-Efficacy.* Building on the central role of intervention skills and defending self-efficacy, prior research has explored education- and training-based approaches as a primary means of empowering bystanders in cyberbullying contexts [52, 53, 95]. These approaches aim to equip individuals with relevant knowledge, strategies, and confidence before they encounter cyberbullying incidents, often through structured curricula, workshops, or simulated practice [119]. For example, Hennessy Garza et al. [53] launched a multi-topic bystander intervention training program for college students called *TAKES ACTION*, aimed at raising students' awareness and knowledge of actively participating in cyberbullying intervention in situations of sexual violence, racism, and high-risk drinking. Hedderich et al. [52] proposed to build a chatbot with the AI that allows teachers to prototype custom dialogue flows and chatbot utterances to enable students to rehearse both desirable and undesirable reactions to cyberbullying in a safe environment, and achieved success initially. Pant et al. [95] proposed and formatively evaluated a classroom-based intervention grounded in social cognitive theory. The program focused on boosting students' defending self-efficacy through media examples, role-plays, and reflective discussions.

Despite these efforts, the effectiveness of education- and training-based interventions remains mixed. Several studies question whether improvements in knowledge or self-reported confidence translate into consistent direct intervention behaviors in real-world cyberbullying situations [20, 87]. This gap motivates the exploration of alternative approaches that can provide in-situ, adaptive, and context-sensitive support, an opportunity that emerging AI-based systems may be particularly well positioned to address.

## 2.3 AI Application in Cyberbullying Mitigation and Its Potential for Direct Intervention

In this part, we first review how AI is currently widely used for cyberbullying mitigation through automated detection (Section 2.3.1); then we point out the potential of AI in promoting defending self-efficacy and skill enhancement, which may serve as a foundation for further promoting the design of direct intervention (Section 2.3.2).

*2.3.1 AI Applications in Cyberbullying Mitigation.* The CSCW community has extensively explored AI-powered cyberbullying detection and content moderation, with growing interest in systems that support human intervention and community response [4, 23, 70, 113]. Specifically, Chang et al. [23] proposed ConvoWizard to provide users with real-time feedback on tensions in online discussions, helping them assess and mitigate the risk of uncivil behavior. Others focus on classifying massive online posts through machine learning algorithms to timely identify potential cyberbullying posts [70, 113]. In addition, AI also plays an important role in actively enforcing social norms [22, 48, 58, 108], which focuses on preventing uncivil, non-compliant, or potentially harmful behavior through design, prompts, or interventions before it actually occurs. Horta Ribeiro et al. [58] proposed "Post Guidance", a community management mechanism that proactively guides user content through rules before posting, rather than waiting to delete or punish after the non-compliant content has been posted. Similarly, Habib et al. [48] proposed and validated a feasible proactive community management approach, using interpretable machine learning to predict the risk of Reddit communities evolving into hate or dangerous groups months in advance, thereby providing a scientific basis for administrators to conduct preventive interventions. Seering et al. [108] designed and deployed the Chillbot tool on Discord, supporting moderators in educating users and mitigating conflicts through anonymous and rapid back-channel feedback, effectively shaping community norms without relying on post deletion or punishment.

These works emphasize the use of automated methods to maintain network order, with AI as the primary focus, rather than aiming to empower bystanders. In this study, we focus more on how AI can actively intervene by enhancing the intervention skills and self-efficacy of bystanders.

*2.3.2 Potential of AI in Promoting Direct Intervention.* While there currently exists minimal research employing AI to directly assist with intervention skills in countering cyberbullying, AI has already been widely applied and supported in other fields to bridge the gaps in interaction ability [13, 65, 103] and self-efficacy enhancement [76, 109, 123], demonstrating the potential of using AI to empower bystanders and thereby promote direct intervention. For example, AI-mediated communication (AIMC) approaches [49] often use AI assistance to improve communication processes by bridging communication skill gaps. Kannan et al. [65] proposed an end-to-end system to automatically generate high-quality, diverse, and scalable email reply suggestions. Boomerang [13] can enhance the users' performance by helping them achieve the appropriate tone in email writing through textual adjustments. Moreover, Sharma et al. [109] used AI in mental health to overcome the inability of peers in writing supportive responses [68, 85], enhancing users' ability and self-efficacy to provide peer support.

Therefore, in this study, we attempt to explore how AI affects bystanders' immediate intervention capabilities and defending self-efficacy in the context of cyberbullying, and impacts the intervention behaviors.

## 3 Method

## 3.1 Overview of Study Design

To investigate how AI-assisted support influences bystanders' direct intervention behaviors in cyberbullying contexts, we adopted a multi-stage mixed-methods research design. Specifically, our methodology consisted of three steps: (1) a pre-study (Section 3.2) to design and validate experimental stimuli, (2) the design and implementation of an AI-assisted intervention system (*EmojiGen*, see Section 3.3 and Section 3.4) and a simulated social media platform (*SnapShare*, see Section 3.5), and (3) a controlled between-subjects experiment evaluating the behavioral and psychological effects of EmojiGen.

The pre-study was conducted prior to the main experiment to ensure that the social media posts and cyberbullying comments used as stimuli were perceived as realistic and comparable across different topics and cyberbullying categories. This step was critical for minimizing potential confounding effects arising from stimulus imbalance. Following the pre-study, we implemented EmojiGen and integrated it into SnapShare, a custom-built social media platform designed to simulate realistic online interactions. Finally, we conducted a between-subjects experiment to examine how EmojiGen affected bystanders' direct intervention behaviors, subjective perceptions, and underlying psychological mechanisms.

## 3.2 Pre-study: Stimuli Design and Validation

*3.2.1 Purpose of the Pre-study.* Although previous research has extensively defined the types of cyberbullying [112] and explored its potential harms [14, 71, 105], these findings remain insufficient to support the design of a rigorous controlled experiment in our research context. Bystanders'

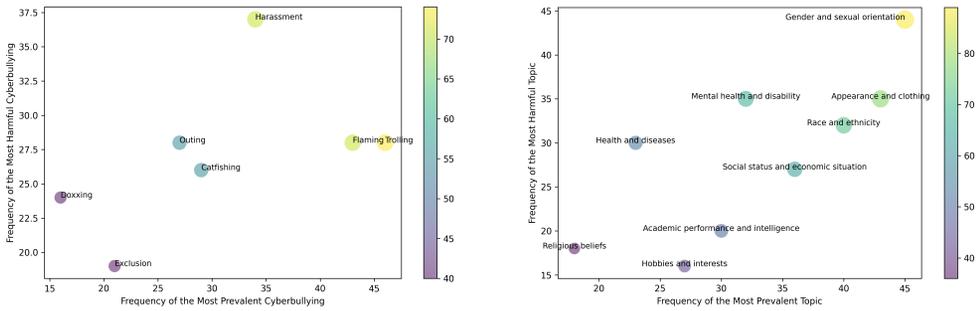

(a) Results for the most prevalent and harmful forms of cy-berbullying.

(b) Results for post topics most likely to provoke cyberbullying and those causing the most severe incidents.

Fig. 1. Voting results from 67 participants for the most prevalent and harmful cyberbullying and post topics most likely to provoke cyberbullying and those causing the most severe incidents. The result suggests that trolling, flaming, and harassment are the three cyberbullying categories that are most prominent in these two indicators. Regarding post-topics, the selected outcomes are gender, race, and appearance.

subjective perception of the experimental stimuli (i.e., social media posts and comments) largely determines whether they choose to intervene [59]. Therefore, before conducting the main experiment, we conducted a pre-study with two objectives. First, we conducted a ***survey*** to identify common cyberbullying categories and post topics that bystanders perceive as prevalent and harmful in real-world social media environments. Second, after carefully designing the experiment materials based on these findings, we sought to ***validate*** that the experimentally constructed posts and cyberbullying comments were perceived as similarly realistic across different topics and categories of cyberbullying (Section 3.2.5). This validation step ensured that subsequent differences observed in the main experiment could be attributed to the intervention rather than artifacts of stimulus design.

*3.2.2 Participants.* We recruited 67 social media users (36 male, 26 female, and 5 identifying as other), aged between 18 and 34 years (M = 23.76, SD = 3.44), through online community postings. All participants reported regular use of social media platforms such as Facebook or Instagram. Participation was voluntary, and attention checks were included to ensure data quality.

*3.2.3 Survey Procedure.* The pre-study was conducted as an online survey consisting of two sequential tasks. Participants were first asked to reflect and identify the most common and harmful forms of cyberbullying they encountered in their daily online experiences. The response options were based on prior literature and expert consultation and included categories [1, 84] such as trolling, harassment, flaming, doxxing, etc. For each option item, we attached the detailed definition to ensure that the participant understood its meaning, for instance, *Trolling (e.g., A posts inflammatory comments to provoke B).* Second, participants were asked to identify post topics that were most likely to provoke cyberbullying and those perceived as causing the most severe harm to victims. Topic options, such as appearance, gender, race, and etc., were drawn from prior studies [16] and refined through discussions with a domain expert in cyberbullying research. An option of the post topic was *appearance and clothing (e.g., A makes fun of B's weight and fashion choices).* Participants completed an attention check before their responses were included in the analysis.

Table 1. Realism ratings for post topics and cyberbullying comment categories. One-way ANOVA results show that realism did not significantly differ across post topics (appearance, gender, race) or across cyberbullying comment categories (harassment, trolling, flaming).

| Condition | Mean (M) | SD |
|---|---|---|
| *Post Topics* | | |
| Appearance | 4.148 | 0.602 |
| Gender | 4.259 | 0.507 |
| Race | 4.241 | 0.594 |
| *Cyberbullying Comment Categories* | | |
| Harassment | 4.047 | 0.442 |
| Trolling | 4.133 | 0.601 |
| Flaming | 4.080 | 0.512 |
| *One-way ANOVA* | | |
| Post topics | $F(2, 78) = 0.295,\ p = 0.745$ | |
| Comment categories | $F(2, 78) = 0.175,\ p = 0.840$ | |

*3.2.4    Survey Result and Stimuli Design.* As shown in Fig. 1(a), **trolling, flaming, and harassment** are three cyberbullying categories that were considered most prevalent and harmful by cyber-by-standers. Regarding post topics (see Fig. 1), **gender, race, and appearance** were selected to be more likely to provoke cyberbullying and cause the most severe outcomes. This result presents the subjective judgments of social media users on cyberbullying in posts and comments, providing empirical evidence for subsequent research design (see Section 3.2.4).

Drawing on real-world examples of cyberbullying from Facebook and Instagram, we crafted nine unique posts spanning three common target topics (appearance, gender, and race). These posts were carefully designed under the supervision of cyberbullying researchers to ensure authenticity and relevance to online harassment experiences. Each post included a set of 1-6 images sourced from publicly available images on Google to enhance their truthiness [91]. These images underwent a rigorous anonymization process to protect individual privacy and uphold ethical research standards while maintaining visual realism. To further diversify the stimuli, three distinct cyberbullying comments were meticulously crafted for each post (a total of 27 comments) to represent the three core cyberbullying categories: trolling, harassment, and flaming. The length, complexity, and emotional tone of these comments were carefully calibrated to reflect typical adolescent online communication patterns observed on platforms such as Facebook and Instagram. Moreover, these cyberbullying comments were interspersed among a larger set of neutral or positive comments beneath each post, simulating a realistic social media feed. The order of appearance of each cyberbullying comment relative to normal comments was randomized to control for any potential order effects. This ensured that participants encountered cyberbullying comments in a realistic and unpredictable manner.

*3.2.5    Realism Evaluation and Results of the Stimuli.* After creating the posts and cyberbullying comments, the realism evaluation survey was conducted, which included another 27 recruited participants (14 male, 10 female, and 3 unwilling to disclose gender), who serve as frequent social media users, providing perceived realism judgments of plausibility and naturalness in social media contexts. All participants reported using social media multiple times a day. The mean age of the participants was 22.385 (*SD* = 4.471). The realism was evaluated with a 5-point scale (1 = Strongly Disagree; 5 = Strongly Agree). For example, *"Do you think this post exists in the real world? <Just* *got this new spaghetti strap dress! It looks amazing, can't wait to wear it to school tomorrow>"*. Each person's average rating of a specific topic post was used to represent the person's rating score on this post topic. The same operations are applied to comment categories.

The evaluated result using one-way ANOVA showed that there was no significant difference in the realism assessment for different post topics ($F(2, 78) = 0.295, p = 0.745$) or different cyberbullying comment categories ($F(2, 78) = 0.175, p = 0.840$). Before conducting the ANOVA, we assessed the assumptions of normality and homogeneity of variances using the Shapiro-Wilk test and Levene's test, respectively. The descriptive statistics for realism are $M = 4.148; SD = 0.602$ for appearance, $M = 4.259; SD = 0.507$ for gender, and $M = 4.241; SD = 0.594$ for race. Regarding the comment categories, the descriptive statistics are $M = 4.047; SD = 0.442$ for harassment, $M = 4.133; SD = 0.601$ for trolling, and $M = 4.080; SD = 0.512$ for flaming. These results demonstrate that our post and comment design has good realism, and the feelings brought to the bystanders by different post topics and cyberbullying categories are relatively balanced, ensuring the materials in the experiment are usable.

## 3.3 Design of EmojiGen

Direct intervention in cyberbullying is often inhibited not only by a lack of moral intent like responsibility [31], but also due to the practical difficulty of composing appropriate responses in real-time [17, 110]. According to cognitive load theory [115], high task complexity may directly suppress action initiation. Self-efficacy theory [41, 127] also suggests that increased cognitive demand may undermine individuals' confidence in their ability to act. Therefore, we posit that lowering the mental effort required to initiate action is a critical enabler of direct intervention. EmojiGen addresses this by transforming the initiation of a comment from a high-effort linguistic task into a low-effort emotional selection.

Importantly, the use of emojis as input is grounded in both behavioral prevalence and communicative efficacy. Emojis have become an ubiquitous form of emotional expression online, offering a compact yet powerful way to convey affect and stance [26, 45, 78, 81]. Their capacity to reduce ambiguity and encode emotional nuance makes them especially suitable for emotionally charged contexts like cyberbullying [66, 104]. By letting users begin with a single emoji that reflects their instinctive response, such as anger, empathy, disapproval, EmojiGen taps into existing expressive habits and reduces initiation friction, while still preserving the richness of emotional intent.

To transform these minimal cues into meaningful social actions, EmojiGen employs an LLM to generate context-aware textual comments that align with both the content of the post and the bystander's selected emotion. LLMs excel at amplifying sparse input into structured, socially appropriate language, making them well-suited to scaffold interventions that are emotionally expressive yet constructive [2, 21]. Additionally, the LLM's tendency toward prosocial tone serves as a natural safeguard against cyberbullying escalation [89], ensuring that even emotionally intense reactions result in supportive or corrective rather than aggressive responses [32, 102].

Together, these three principles, lowered cognitive cost, emotion-driven initiation, and LLM-powered expansion, form the foundation of EmojiGen's design.

## 3.4 Implementation of EmojiGen

*3.4.1 Emoji Selection.* Since this is the first study to combine emojis with AI to promote cyberbullying intervention, we decided to limit the selection of emojis as some emojis can convey complex and ambiguous emotions. After discussing with the domain expert in cyberbullying, we finally selected the emojis that clearly express emotions, which were categorized into two groups: resisting the perpetrator and comforting the victim.

*Resisting the Perpetrator:* The following emojis are chosen to express defiance towards the perpetrator: 😒 (displeased), 😠 (angry), 🤬 (cursing), 😑 (unimpressed), 🙄 (meh), 🙄 (eye roll), 🤐 (zip it), 🤡 (clown), 😾 (angry cat), 👊 (fist), 🤛 (rightward fist), 👉 (you), 🙅 (no), 🤷 (who cares), 🙉 (not listening).

*Comforting the Victim:* To express support and comfort towards the victim, the following emojis are selected: 😉 (winking face), 😌 (relieved face), 🤗 (hugging face), 🫶 (heart hands), 🤝 (handshake), 😼 (smirking cat), 😻 (heart eyes cat).

*3.4.2 Comment Generation with EmojiGen.* The generation of EmojiGen depends on the bystander's browsing context and the selected emojis. When the bystander clicks on the 'Comment' or 'Emoji-Gen' button under a post, the context would include only the post, while the bystander clicks on 'Comment' or 'EmojiGen' under someone's comment, the context would involve both the current post and the targeted comment.

To support context-sensitive and emotionally grounded comment generation, we designed two prompt templates tailored to the bystander's interaction context: one for replying to a post and the other for replying to a specific comment (see Table 2). These prompts were carefully and iteratively crafted to ensure its stable output. Additionally, to ensure the generated content resonates with the intended user group, typically teenagers or young adults, we instruct the model to adopt a peer-like tone. Furthermore, the model was requested to avoid excessive politeness or verbosity, encouraging concise and direct expressions that match the emotional stance.

Fig. 2 demonstrates a generation process. When the bystander was browsing the post, he found a trolling comment under this post in the appearance topic. He then clicked on the 'EmojiGen' button and selected a 🙉 in the pop-up emoji panel. The system inferred his intention to resist from the selected emoji and then generated a resisting response based on the given browsing context: "🙉 *Dude, no one wants to hear that outdated nonsense.*"

## 3.5 SnapShare - Development of the Experimental Platform and Authenticity Control

To investigate the effectiveness of EmojiGen in a realistic social media environment, we developed an online social media platform, SnapShare, incorporating the core functionalities of other well-known social media platforms such as user registration, editing profiles, creating new posts, browsing published posts, and acting (flag, like, reply to), etc. EmojiGen was enabled on this platform, depending on the experimental setup.

We leveraged Vue.js[1] with javascript[2] for front-end development, and took Django[3] with Python[4] to build the back-end server. This application was installed on Amazon Web Services[5] for public access during data collection. This experimental platform can only be accessed from a PC and is not compatible with mobile phones. The common features *'friending'*, *'following'*, and *'private messaging'* other users were excluded, as these were unnecessary to evaluate our design. It orchestrated concurrent simulations for each participant to ensure that each participant experienced a similar yet authentic social environment, with differences arising only from the experimental conditions and their own activities.

To ensure participants experienced a highly authentic and immersive environment, we approached platform development in two distinct phases: focusing first on interface authenticity

---

Table 2. Prompt templates used in EmojiGen for generating comments. The system dynamically fills in user-specific and context-specific fields (e.g., my_username, post_content, selected_emoji, and timestamp).

| Context Type | Prompt Template |
|---|---|
| **Post-level** | ```[POST publisher]: {post_username}```<br>```[POST]: {post_content}```<br>```I am a reader of this [POST] and my [USERNAME]: {my_username}```<br>```My [Emoji] is {selected_emoji}```<br>```The content you generate should:```<br>```- According to the context, guess the meaning my emoji wants to express```<br>```- Follow the meaning expressed by the emoji I selected```<br>```- Reply to young audiences using vocabulary and expressions they are familiar with```<br>```- Not be long and repeat```<br>```Note: You only need to provide the content you think they want to comment on, and do not generate any other nonsense, your output does not need to be in quotes``` |
| **Comment-level** | ```[POST publisher]: {post_username}```<br>```[POST]: {post_content}```<br>```[Commenter] {commenter_username} comment: "{comment_content}" to this [POST]```<br>```I am a reader of this [POST] and my [USERNAME]: {my_username}```<br>```My [Emoji] to [{commenter_username}] is: {selected_emoji}```<br>```The content you generate should:```<br>```- According to the context, guess the meaning my emoji wants to express```<br>```- Follow the meaning expressed by the emoji I selected```<br>```- Reply to young audiences using vocabulary and expressions they are familiar with```<br>```- Not be long and repeat```<br>```Note: You only need to provide the content you think they want to comment on, and do not generate any other nonsense, your output does not need to be in quotes``` |

and then on content authenticity. SnapShare's basic function design and layout were based on the EatSnap.Love platform, which has been assessed in another cyberbullying intervention study [37], is considered authentic enough. Afterward, we replicated posts and their associated comments from well-known social media platforms such as Facebook and Instagram. We then created 30 virtual accounts by adapting and modifying the nicknames of users from these platforms to post relevant content or make comments as if they were actual users. The pilot test with 15 participants was then conducted to evaluate the interface's usability and credibility. All of them were recruited from the first-author's university, including five PhD students, seven Master's students, and three Bachelor's students. Their mean age was 23.07 years ($SD$ = 2.60). They all have extensive experience using social media. Based on their feedback, we refined the platform interface.

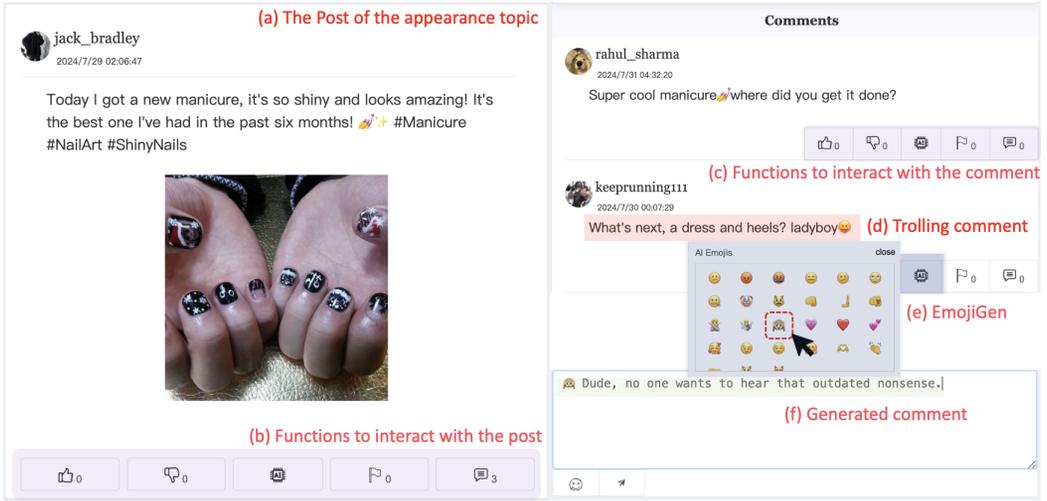

Fig. 2. The screenshot of SnapShare that allows the use of EmojiGen. On the left side (a) is a post of the appearance topic, below the post is the (b) button group that allows participants to interact with the current post. The buttons from left to right are liking, disliking, EmojiGen, flagging, and commenting. On the right side, there is the comment display board where participants can also interact with each comment through its (c) button group using the same five interaction methods. (d) is a trolling cyberbullying comment. The participant is using (e) EmojiGen to resist. Specifically, the participant selected a 🙎 on the pop-up emoji panel, and then the system (f) suggested a resisting response according to the current browsing context. The generated content is: 🙎 Dude, no one wants to hear that outdated nonsense.

Subsequently, we embedded the series of posts and comments (refer to Section 3.2.5) that have been well-designed and underwent a realism evaluation. For the formal experiment (N = 90), when participants completed their task and before they were informed about the actual purpose of this study, we asked them to share feelings. Almost all participants reported that SnapShare felt realistic, and none expressed doubts about the authenticity of the posts and comments. Only three participants speculated that the study platform might be related to cyberbullying.

## 3.6 Participants

We recruited 90 participants (54 females and 36 males) by posting announcements in local online communities. We first posted recruitment information in the local Telegram community, emphasizing that participants need to have extensive social media experience. In the announcement, we used a preliminary survey to screen and select particular participant groups. We targeted a young demographic to avoid potential biases in understanding and behavior caused by age differences [6], and the mean age of the participants finally recruited was 27.112 (SD = 5.144). In addition, we also included a screening item regarding the willingness to intervene: *"When you encounter someone being cyberbullied on social media, your usual response is"* with options: *"Direct intervention (e.g., posting comments to stop the cyberbully or to comfort the victim), Indirect intervention (e.g., disliking, flagging, or reporting the content), Both direct and indirect interventions, and Ignoring"*. All participants were selected because they reported that they would ignore (N = 32) or intervene in cyberbullying indirectly (N = 58) in daily life. This screening aimed to select participants with lower direct intervention in their past experience to avoid experimental bias caused by initial intervention willingness. They would receive a reward of $6 for system interaction and survey completion. The participants' education levels were distributed as follows: 21 had completed high school, 58 held a bachelor's degree, and 11 had a master's degree. The study was carried out following a rigorous procedure approved by the Institutional Review Board (IRB) of the author's university.

### 3.7 Experimental Settings

During the main task, each participant browsed a simulated social media feed consisting of nine posts. Each topic (gender, race, and appearance) has three variate posts. Each post contained a mixture of several neutral and positive comments, as well as one cyberbullying comment, mimicking a realistic online environment. Participants were not instructed to intervene and were free to interact with or ignore any content.

This experiment used a between-subjects design, with one group ($C_{EG}$, N = 45) having access to EmojiGen while the other does not ($C_{NE}$, N = 45). Data collection was spread over three days, with each day having different order settings. Each participant only participated in the experiment once. When conducting the experiment, we avoided order effects on different **post topics** by controlling the visibility sequence (see Table 3). In Table 3, vertically, the topic order on day 1 is *appearance, gender,* and *race*, indicating that participants on day 1 would first see 3 posts of *appearance* (A1, A2, A3) when landing on the system. Then, as they scrolled down, they would gradually see posts of *gender* (G1, G2, G3) and *race* (R1, R2, R3).

Table 3 also reports all variants of combination between **cyberbullying comments** and **posts**. This arrangement ensured that each post on the same topic could feature all three types of cyberbullying (*trolling, flaming,* and *harassment*), and prevent co-occurrence between any post and cyberbullying comments (see Table 8 in Appendix A). For instance, on day 1, the post A1 co-occurred with *Trolling*, while on day 2, it was paired with *Flaming*. Table 3 omits showing other neutral or positive comments, which are identical for all participants. For clarity, in Table 4, we present all three posts and their comments under the theme of ***appearance*** on day 3. Additionally, Table 6 and Table 7 in Appendix A shows the posts and comments under the other two topics (*gender* and *race*) on day 3. Table 8 in Appendix A represents the combination of all posts and cyberbullying comments.

### 3.8 Procedure

To establish a socially meaningful interaction context, participants were told that SnapShare was a beta version of a new social networking platform, and that the posts and comments were created by other users participating in the beta test. Participants were also informed that the study was part of an ongoing research effort and that they might be contacted for follow-up studies. This framing was intended to encourage participants to perceive their actions as socially situated and potentially consequential, approximating early-stage interactions on emerging social media platforms.

Through the pre-survey described in Section 3.6, we selected suitable participants and collected their demographic and availability information. Then, participants received a link to access the platform at the scheduled time. They registered with the same email address used for the pre-survey, and chose a public nickname for the platform. Upon logging in, they were required to peruse the comprehensive instructions to grasp the site's functionality, their responsibilities, and the code of conduct for engaging in the test [37], shown in Fig. 3. Specifically, they were first informed about the (A) available functionality on the platform, including like, dislike, flag, and comment features; then, a page with detailed instructions on how to use EmojiGen would be (B1) accessible or (B2) not accessible, depending on their assigned experimental condition. If displayed, the (C) guidance of how to use EmojiGen would display as follows. Afterward, (D) community guidelines were shown, highlighting the participants' duty to help maintain a positive and cooperative online environment. Finally, participants were asked to confirm their agreement. Subsequently, the participant engaged in the 'Square' page that includes pre-defined posts and comments. The total interaction time was about 10 minutes. After the participant completed the interaction, they clicked the 'Finish' button to process the post-survey and reflect on their experience of using this platform, as well as their attitude towards cyberbullying.

Table 3. Experimental arrangement of posts and cyberbullying comments displaying over three-day experiments. Each participant only participated in the experiment for one day. A1, A2, and A3 represent three different posts under the **A**ppearance topic, with similar representations for other topics. A1 (Trolling) means that the type of cyberbullying comment that occurs under the post A1 is Trolling. Under each post (e.g., A1), there are other neutral and positive comments to simulate real online interactions, rather than just having cyberbullying comments.

| Day (participants) | Topic Order ↓ | Post Order (comment type) → | | |
|---|---|---|---|---|
| Day 1 (15 $C_{NE}$ / 15 $C_{EG}$) | Appearance | A1 (Trolling) | A2 (Flaming) | A3 (Harassment) |
| | Gender | G1 (Trolling) | G2 (Flaming) | G3 (Harassment) |
| | Race | R1 (Trolling) | R2 (Flaming) | R3 (Harassment) |
| Day 2 (15 $C_{NE}$ / 15 $C_{EG}$) | Gender | G1 (Flaming) | G2 (Harassment) | G3 (Trolling) |
| | Race | R1 (Flaming) | R2 (Harassment) | R3 (Trolling) |
| | Appearance | A1 (Flaming) | A2 (Harassment) | A3 (Trolling) |
| Day 3 (15 $C_{NE}$ / 15 $C_{EG}$) | Race | R1 (Harassment) | R2 (Trolling) | R3 (Flaming) |
| | **Appearance** | **A1 (Harassment)** | **A2 (Trolling)** | **A3 (Flaming)** |
| | Gender | G1 (Harassment) | G2 (Trolling) | G3 (Flaming) |

## 3.9 Measurements

To answer our research questions, we collected quantitative data from system logs and the questionnaire, as well as qualitative data from open-ended responses and interviews of the participants. Unless otherwise noted, all self-report survey measures were calculated as the mean score across their respective items to represent participants' overall responses. All multi-item perceptual measures were examined using exploratory factor analysis (EFA) to assess their underlying factor structure.

### 3.9.1 Direct Engagement Frequency.
This indicator refers to how many times one bystander comments in social media interactions, including *(1) post engagement frequency* and *(2) comment engagement frequency* according to the browsing context when this engagement occurs. Specifically, if the participant comments under a certain post, then this interaction is accounted for post engagement, while if they interact under an existing comment of another user, it is counted as a comment engagement.

### 3.9.2 Direct Cyberbullying Intervention Frequency.
In line with the definitions of direct and indirect intervention provided in [38], direct intervention frequency refers to how many times each participant comments on a "being cyberbullied" post to support victims, called the *(1) direct support frequency*, or directly on an existing cyberbullying comment to resist perpetrators, which is called the *(2) direct resisting frequency*.

All direct supportive behaviors were chosen from the post-engagement records, while all direct resisting behaviors were gathered from the comment engagement records. This process was completed, including independent annotation by the first and second authors, who then discussed any disputed comments until consensus was achieved.

Table 4. **Topic 1: Appearance-related posts and comments.** This table presents posts centered on physical appearance and body-related self-presentation. Each post includes one appearance-focused cyberbullying comment embedded among neutral or supportive comments to simulate realistic social media contexts. The **color** is used to mark the posts and cyberbullying comments to indicate the correspondence of the content in Table 3.

| User | Time | Content (Text-only version, omitting emojis and images) |
|------|------|---------------------------------------------------------|
| **Post A1 (Harassment)** | | |
| Amy_heartthief | 2024/7/13 21:09 | **[POST] Just got this new spaghetti strap dress! It looks amazing, can't wait to wear it to school tomorrow #NewOutfit #Fashion #SchoolStyle** |
| yuki_harada | 2024/7/16 20:33 | [Comment] Can't wait to see you in it tomorrow |
| emma_grace | 2024/7/16 15:32 | [Comment] Wow, that dress is stunning! You're going to look great in it |
| bella_smith | 2024/7/16 06:07 | [Comment] Nice choice! Simple yet elegant |
| liam_anderson | 2024/7/15 22:24 | [Comment - Harassment] Wow, you're way too young to be dressing like that. Trying to seduce men or what? |
| ananya_sharma | 2024/7/14 18:22 | [Comment] Love it! Where did you get it from? |
| **Post A2 (Trolling)** | | |
| ashley_star | 2024/7/8 04:02 | **[POST] Is there any outfit that looks good on a girl who is 160cm and 80kg #FashionStruggles #BodyPositivity #NeedHelp** |
| sakura_matsui | 2024/7/15 05:45 | [Comment] Try some flowy tops and high-waisted jeans, they can be really flattering |
| meilin13 | 2024/7/8 07:38 | [Comment] It's not about the clothes, it's about confidence. Own it! |
| arjun_singh | 2024/7/8 05:08 | [Comment - Trolling] Wow, you're really asking for fashion tips? Being this fat disqualifies you from talking about "beauty". |
| **Post A3 (Flaming)** | | |
| JamesLover23 | 2024/7/23 23:24 | **[POST] Just had my first gym session! Hoping to get as strong as LeBron James in a year #GymLife #FitnessGoals** |
| mike_the_great | 2024/7/26 19:06 | [Comment] Great start, man! Keep pushing and you'll get there |
| tommy_jameson | 2024/7/25 23:21 | [Comment] Way to go! Remember, consistency is key |
| lily_white | 2024/7/25 20:21 | [Comment - Flaming] Such a twig doesn't deserve to be called a man, more like a tiny boy. |

*3.9.3 Defending Self-efficacy.* Defending self-efficacy [118] refers to the belief that individuals can intervene and protect victims successfully. In line with [28], the Clark and Bussey's defending self-efficacy scale was adapted to measure self-efficacy to engage in common defending strategies employed in cyberbullying episodes. Participants rated on a 7-point Likert scale (1 = No Confidence; 7 = Very Confident) how confident they feel for four specific scenarios, including *"If see cyberbullying on SnapShare, I will tell the bully to stop," "If I see cyberbullying on SnapShare, I will provide comfort to the people who are being cyberbullied", "If I see cyberbullying on SnapShare, I will tell the bully that what they are doing is not okay"*, and *"If I see cyberbullying on SnapShare, I will defend the people who are being cyberbullied by being mean to the bully"*. This scale showed good internal consistency ($\alpha = 0.85$), and EFA supported a unidimensional structure, with standardized factor loadings ranging from 0.63 to 0.89. Descriptive statistics for the composite defending self-efficacy score are $M = 3.500, SD = 1.423$.

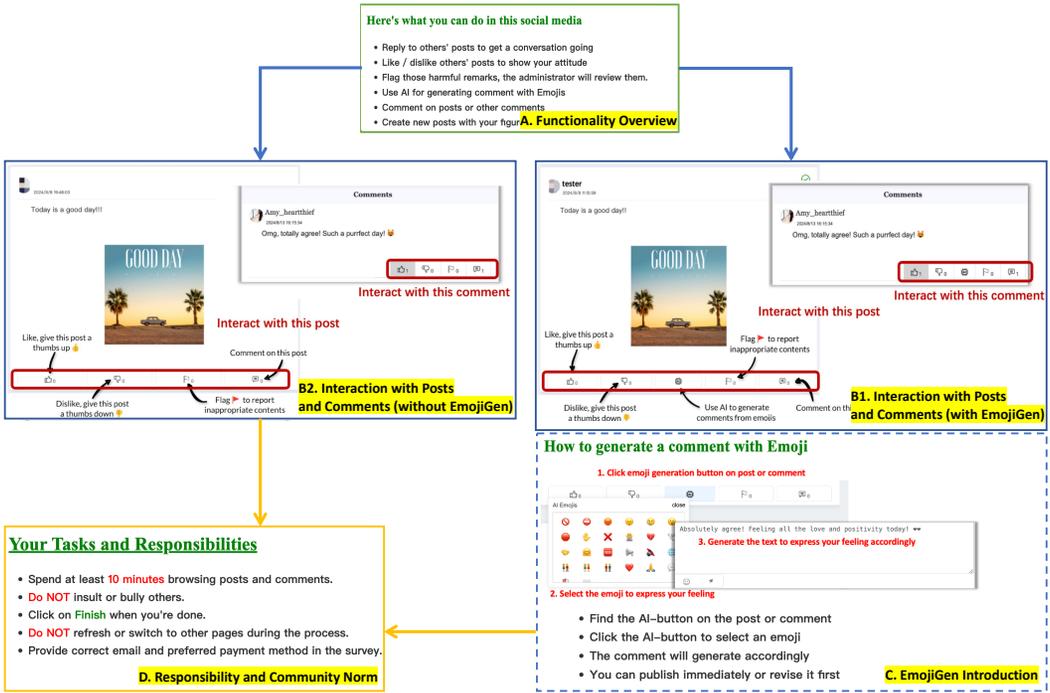

Fig. 3. Guidance provided to participants during the SnapShare testing process, consisting of four steps (A–D). Step A offers a functionality overview to brief participants on the available features. Step B introduces specific interaction behaviors, with two interface versions (B1/B2) presented based on the experimental condition: B1 includes introductions to liking, disliking, flagging, and EmojiGen, while B2 excludes the EmojiGen feature. If EmojiGen is available, participants are shown Step C, which provides a dedicated introduction to EmojiGen. Finally, Step D outlines participant responsibilities and a code of conduct, which participants must agree to before starting the experiment.

*3.9.4 Know How to Help.* In this study, know how to help refers to the extent to which participants possess the knowledge and skills to intervene or provide assistance to victims of cyberbullying [92]. We adopted a measurement of three items that were separated from [92] and rated on a 7-point scale (1 = Strongly Disagree; 7 = Strongly Agree). The questions contain *"I have the skills to make a comment to support those who are being cyberbullied on SnapShare"*, *"I know what to comment to get someone to stop cyberbullying someone else on SnapShare"*, and *"I can make a comment to get someone out of a situation where he or she is being cyberbullied on SnapShare"*. This scale demonstrated good reliability ($\alpha = 0.873$), and EFA supported a unidimensional structure, with standardized factor loadings ranging from 0.78 to 0.88. Descriptive statistics for the composite know-how-to-help score are $M = 4.204$, $SD = 1.437$.

*3.9.5 Workload.* In this study, the workload refers to the effort they need to put into commenting on social media interactions. Subjective Workload Assessment Technique (SWAT) [126], which was used to assess the perceived mental workload of the participants during a specific task, was modified to assess the workload. By setting the logic of the survey, only those who make comments during the experiment need to report their perception of the workload. Specifically, it focuses on three dimensions: time load, mental effort load, and psychological stress load. Each dimension is rated on a scale from 1 (Lowest Load) to 7 (Highest load). The questions contain: *"I think it needs* *lots of my time when making comments on SnapShare", "I think it needs lots of my attention and cognitive effort when making comments on SnapShare*, and *"I think it needs lots of my psychological stress (feel tense, frustrated, or emotionally strained) when making comments on SnapShare".* This scale showed good internal consistency ($\alpha = 0.870$), and EFA supported a unidimensional structure, with standardized factor loadings ranging from 0.76 to 0.97. Descriptive statistics for the composite workload score are $M = 3.267, SD = 1.493$.

*3.9.6 Anxiety.* In this study, anxiety refers to the stress regarding one's ability and skills to intervene when faced with cyberbullying and the intention to engage in direct intervention. For measurement, we devised an anxiety scale adapted from the State-Trait Anxiety Inventory (STAI) [114]. The scale consists of four items: *"I feel anxious when preparing to comment on a cyberbullying incident on SnapShare," "I feel pressure that I might say something wrong when confronting cyberbullying on SnapShare," "I feel anxious that my words might not be strong enough when addressing cyberbullying on SnapShare,"* and *"I worry that my comment will receive negative feedback when I try to stop cyberbullying on SnapShare."* Respondents rated their agreement with these statements on a Likert scale ranging from 1 (Strongly Disagree) to 7 (Strongly Agree). This scale demonstrated good internal consistency ($\alpha = 0.871$), and EFA supported a unidimensional structure, with standardized factor loadings ranging from 0.76 to 0.89. Descriptive statistics for the composite anxiety score are $M = 4.564, SD = 1.366$.

*3.9.7 Personal Responsibility.* In this research, personal responsibility indicates how much participants feel obligated to assist when they observe cyberbullying [37]. We adopted a three-item measure of personal responsibility for cyberbullying from Pozzoli and Gini [101], including *"I feel personally responsible to intervene and assist in resolving cyberbullying incidents on SnapShare", "If I am not the one cyberbullying others, it is still my responsibility to try to stop it on SnapShare"*, and *"I believe that my actions can help reduce cyberbullying on SnapShare".* All items were measured on 7-point Likert scales (1 = Strongly Disagree; 7 = Strongly Agree). This scale showed good internal consistency ($\alpha = 0.832$), and EFA supported a unidimensional structure, with standardized factor loadings ranging from 0.67 to 0.92. Descriptive statistics for the composite personal responsibility score are $M = 4.307, SD = 1.325$.

*3.9.8 Qualitative Data Collection.* To better understand the subjective experience of participants when using EmojiGen for AI utility, we collected their specific insights through the question in the post-survey: *"How do you think the effectiveness of AI generation in helping you deal with cyberbullying or making comment on SnapShare."* Afterward, we asked the user to review *"whether and how they have modified comments on AI on SnapShare and explain in detail the reasons for doing so"* to explore the limitations of AI-generated content and design insights.

## 3.10 Data Analysis

To address our research questions, we conducted a series of quantitative and qualitative analyses, organized by research question. All quantitative analyses were conducted using JASP (v0.18.3)[6].

*3.10.1 Analysis for RQ1: Direct Intervention Behaviors.* To examine whether EmojiGen influenced bystanders' direct intervention behavior (RQ1), we analyzed two primary behavioral outcomes derived from system logs: *direct support frequency* (comments posted to support victims) and *direct resisting frequency* (comments posted to confront perpetrators). These measures were defined and annotated as described in Section 3.9.2.

---

[6]https://jasp-stats.org/

We first compared the frequencies of direct support and direct resisting behaviors between the EmojiGen condition ($C_{EG}$) and the control condition ($C_{NE}$) using independent-samples t-tests. Prior to hypothesis testing, we assessed the assumptions of normality and homogeneity of variances using the Shapiro-Wilk test and Levene's test, respectively. When violations of homogeneity were detected, Welch's t-test was applied.

To further explore whether intervention behavior varied across different post topics (appearance, gender, and race) and cyberbullying categories (trolling, flaming, and harassment), we conducted one-way ANOVAs, followed by post-hoc Tukey HSD tests.

*3.10.2 Analysis for RQ2: Perceptual Measures.* To investigate how EmojiGen affected bystanders' perceptions related to cyberbullying intervention (RQ2), we analyzed participants' self-reported measures, including defending self-efficacy, sense of knowing how to help, workload, anxiety, and personal responsibility.

We compared these perceptual outcomes between the EmojiGen condition ($C_{EG}$) and the control condition ($C_{NE}$) using independent-samples t-tests. Assumptions of normality and homogeneity of variance were evaluated prior to analysis, and Welch's t-test was applied when assumptions were not met.

*3.10.3 Analysis for RQ3: Mechanism Exploration via Regression and Mediation.* To examine the mechanism through which EmojiGen influenced bystanders' direct intervention behaviors (RQ3), we conducted linear regression and mediation analysis. All analyses for RQ3 were conducted at the participant level. Although participants could engage in multiple interactions during the task, all behavioral measures (i.e., *direct support frequency* and *direct resisting frequency*) were aggregated per participant based on system logs. Likewise, all perceptual variables (e.g., *defending self-efficacy, know how to help, anxiety, workload,* and *personal responsibility*) were collected once per participant via the post-task questionnaire.

We first applied linear regression with *forward* selection to identify key predictors of direct supporting and direct resisting behaviors (see Section 4.4.1 and Section 4.4.2). This exploratory step allowed us to identify perceptual variables that were most strongly associated with intervention behaviors.

Based on the regression results and prior theoretical work, we then specified a structural equation model (SEM) to examine the indirect pathways through which EmojiGen usage affected direct intervention behavior via intermediate perceptual factors. Importantly, this SEM was estimated using data from all participants (N = 90), with *EmojiGen usage* modeled as a binary between-subjects variable (use EmojiGen or not), rather than restricting the analysis to participants in the EmojiGen condition alone. This approach allowed us to examine how the experimental manipulation influenced intervention behavior through changes in bystanders' perceptions across the full sample.

Given the modest sample size, we used bootstrap resampling (1,000 samples) to estimate standard errors and 95% confidence intervals for indirect effects. The SEM results should therefore be interpreted as theory-informed and exploratory, aimed at elucidating plausible mechanisms rather than providing confirmatory causal evidence.

*3.10.4 Analysis of Qualitative Data.* To analyze participants' open-ended survey responses and interview reflections, we conducted a reflexive thematic analysis [15]. This approach was chosen to support an interpretive understanding of participants' experiences with EmojiGen, rather than to categorize responses or quantify theme frequencies. The analysis was primarily inductive and theoretically informed by prior work on intervention skills, defending self-efficacy, cognitive load, and anxiety. Following Braun and Clarke's analytic framework, we familiarized ourselves with the data, generated initial codes, and iteratively developed and refined themes through repeated engagement with the dataset. In line with reflexive thematic analysis, the first author led the analysis, with themes refined through regular discussions among the research team to support reflexivity and analytic coherence. Rather than aiming for inter-rater reliability, these discussions focused on strengthening interpretation and theoretical alignment. The resulting themes were used to complement the quantitative findings by explaining how and why EmojiGen shaped bystanders' perceptions and behaviors, and to inform the design implications.

## 4 Result

This section presents the results of our investigation into the impact of EmojiGen on direct intervention behaviors of people (RQ1), their perceptions (RQ2), and the mediation effects (RQ3). To facilitate the presentation of results, in the following description, $C_{EG}$ is used uniformly to represent the group using EmojiGen, while $C_{NE}$ is for the group where EmojiGen is not allowed.

### 4.1 Overall Statistic of Bystanders' Engagement

This section encompasses all records of direct bystander engagement, such as their direct interactions with posts and comments from other users. Additionally, it offers an in-depth overview of how bystanders utilized EmojiGen during these interactions (see Fig. 4).

*4.1.1 Post Engagement.* We counted the post engagement of the participants, totaling 337 interactions. Of these, 123 (36.4%) instances are from $C_{NE}$ and 214 (63.6%) instances are from the $C_{EG}$ group. Further component analysis shows that out of the 214 instances, 18 (8.4%) interactions were entirely composed by the participants without EmojiGen assistance, while 196 (91.6%) interactions involved at least one use of EmojiGen, where 138 (70.4%) were published without modification, while 18 (9.2%) were modified before being shared. However, 40 (20.4%) interactions that were initially generated by EmojiGen were abandoned and finally written by the participants themselves (see Fig. 4(a)).

*4.1.2 Comment Engagement.* In terms of comment engagement, 181 interactions were recorded in total. Of these, 36 (19.9%) instances were in the $C_{NE}$ group, and 145 (80.1%) instances were in the $C_{EG}$ group. Detailed analysis reveals that of the 145 instances, 5 (3.4%) interactions were crafted solely by participants without any help from EmojiGen. The remaining 140 (96.6%) interactions involved EmojiGen usage at least once, of which 96 (68.6%) posted without changes and 14 (10%) modified before sharing. Nevertheless, 30 (21.4%) interactions initially generated by EmojiGen were eventually discarded and rewritten by the participants themselves (see Fig. 4(b)).

The results suggest that **EmojiGen influenced how often participants engaged in both posting and commenting**. A majority of participants used AI-generated content directly, whereas others decided to edit or rewrite the content before posting.

### 4.2 EmojiGen Promotes Bystanders' Direct Cyberbullying Interventions (RQ1)

All direct support and resisting behaviors are gathered from the posts and comments engagement records (see Section 3.9.2). Section 3.10.1 gives more details on data analysis. **The results show that using EmojiGen increased the frequency of direct intervention, including both direct resisting and direct support.**

*4.2.1 Direct Support Frequency.* Out of all 337 post engagements, 326 (96.7%) were recognized to provide supportive comments to the victim. T-test results further showed that participants in the group $C_{EG}$ ($M = 4.896; SD = 2.897$) performed a significantly higher frequency of individual direct support ($t(88) = 4.996, p < 0.001$), compared to group $C_{NE}$ ($M = 2.116; SD = 2.342$), suggesting that **EmojiGen effectively promotes bystanders to support the victim in a direct way.**

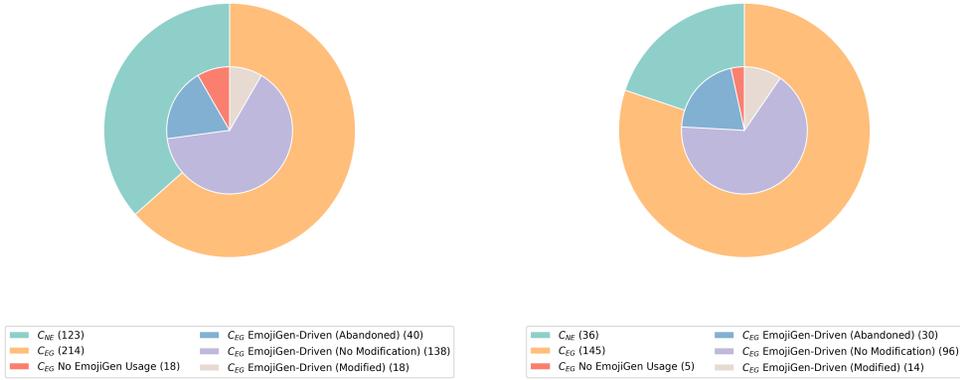

(a) Nested pie chart illustrating the distribution of post engagement.
(b) Nested pie chart illustrating the distribution of comment engagement.

Fig. 4. Nested pie charts illustrating the distribution of (a) post and (b) comment engagement. Note that the inner and outer rings are not proportionally aligned and should be interpreted independently. The outer ring shows the proportion of the engagement frequency between the $C_{EG}$ and $C_{NE}$ groups. The inner ring represents a within-condition breakdown specific to the $C_{EG}$ group only, depicting the number of engagements where EmojiGen-driven content was adopted without modification, adopted with modifications, ultimately abandoned and with no EmojiGen usage.

Furthermore, the following one-way ANOVA showed that the **topic of the post** had a significant effect on the direct support frequency, $F(2, 207) = 4.386, p = 0.014$. The post-hoc analysis using Tukey HSD test revealed that there was a significant difference between **appearance** and **race** ($MD = -0.543, p = 0.011$), while there was no significant difference between **appearance** and **gender** ($MD = -0.186, p = 0.580$), nor between **gender** and **race** ($MD = -0.357, p = 0.136$). The results suggest that in addition to using EmojiGen, the post topics influence participants' supportive behavior, with race-related posts being particularly likely to elicit support for victims of cyberbullying.

*4.2.2 Direct Resisting Frequency.* Out of the 181 comment engagement records, 114 (63.0%) were produced to resist the perpetrators. The Welch's t-test results revealed a significant difference ($t(52.898) = 4.021, p < 0.001$) in the personal direct resisting frequency between participants in the $C_{EG}$ group ($M = 2.125; SD = 3.085$) and the $C_{NE}$ group ($M = 0.279; SD = 0.734$), which means that **EmojiGen also effectively promotes the bystanders to take actions to resist perpetrators.**

The following one-way ANOVA showed that EmojiGen did not have a significant effect ($F(2, 87) = 0.621, p = 0.54$) on the direct resisting frequency among cyberbullying categories, such as flaming, trolling, harassment. This suggests that **EmojiGen's effect on promoting resistance does not vary based on the type of cyberbullying encountered by bystanders.**

## 4.3 The Impact of EmojiGen on Bystander Perceptions (RQ2)

In this section, we present the effects of EmojiGen on participants' subjective perceptions across five measures (see Fig. 5). Section 3.10.2 gives more details on data analysis. Except for **personal responsibility**, all other indicators showed significant differences due to the use of EmojiGen, **indicating EmojiGen's influence on the bystanders' perceptions when they encountered cyberbullying behaviors.**

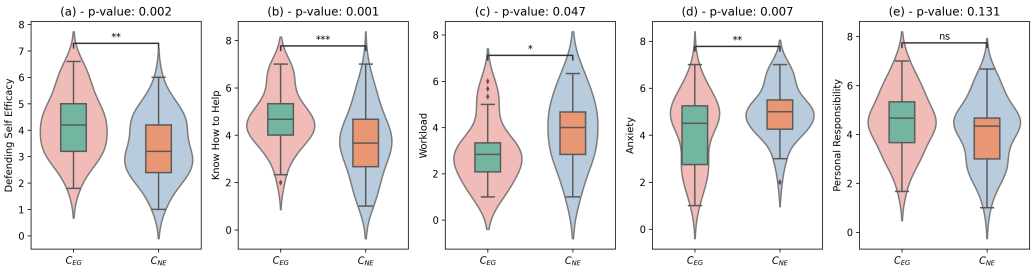

Fig. 5. Comparison of five subjective measures between bystanders in the $C_{EG}$ and $C_{NE}$ groups: (a) Defending self-efficacy, (b) Sense of knowing how to help, (c) Workload, (d) Anxiety, (e) Personal responsibility. The violin plots represent the distribution of scores, with overlaid boxplots showing the median and interquartile ranges. Statistical significance is denoted by p-value annotations: ns ≥ 0.05, * < 0.05, ** < 0.01, *** < 0.001.

(1) **Defending Self-Efficacy.** A significant difference was observed between the $C_{EG}$ and $C_{NE}$ groups in defending self-efficacy, as indicated by the results of the Student's t-test ($t(88) = 3.268, p = 0.002$). Participants in the $C_{EG}$ group ($M = 4.191; SD = 1.207$) exhibited greater confidence in their ability to intervene in cyberbullying compared to those in the $C_{NE}$ group ($M = 3.369; SD = 1.180$). This finding suggests that **EmojiGen usage enhances bystanders' confidence in their ability to effectively intervene in cyberbullying situations.**

(2) **Know How to Help.** Participants using EmojiGen reported significantly higher perceptions of knowing how to intervene ($t(88) = 3.505, p < 0.001$). The mean score for the $C_{EG}$ group was $M = 4.704$ ($SD = 1.229$), while the $C_{NE}$ group scored lower, with $M = 3.704$ ($SD = 1.467$). This result indicates that **EmojiGen helps bystanders feel more prepared to intervene when witnessing cyberbullying.**

(3) **Workload.** The t-test results revealed that participants in the $C_{EG}$ group experienced significantly lower perceived workload when generating comments ($t(43) = -2.047, p = 0.047$). Specifically, the $C_{EG}$ group reported a mean workload of $M = 2.956$ ($SD = 1.386$), compared to $M = 3.889$ ($SD = 1.552$) in the $C_{NE}$ group. This suggests that **EmojiGen reduces the cognitive effort required by bystanders to create intervention comments, making the process easier and less burdensome.**

(4) **Anxiety.** The Welch t-test results showed a significant reduction in anxiety among participants in the $C_{EG}$ group ($t(81.263) = -2.738, p = 0.008$). Anxiety levels were lower in the $C_{EG}$ group ($M = 4.183; SD = 1.497$) compared to the $C_{NE}$ group ($M = 4.944; SD = 1.113$). This finding implies that **EmojiGen helps reduce the emotional stress associated with intervening in cyberbullying situations.**

(5) **Personal Responsibility.** The t-test showed that the difference in personal responsibility was not statistically significant due to the usage of EmojiGen ($t(88) = 1.523, p = 0.131$), participants in the $C_{EG}$ group ($M = 4.519; SD = 1.280$) had slightly a higher mean score than those in the $C_{NE}$ group ($M = 4.096; SD = 1.350$).

## 4.4 Mechanisms of EmojiGen's Influence on Direct Intervention (RQ3)

### 4.4.1 Mechanism Analysis for Direct Resisting Behaviors.
As detailed in Section 3.10.3, we sequentially conducted linear regression and SEM analysis.

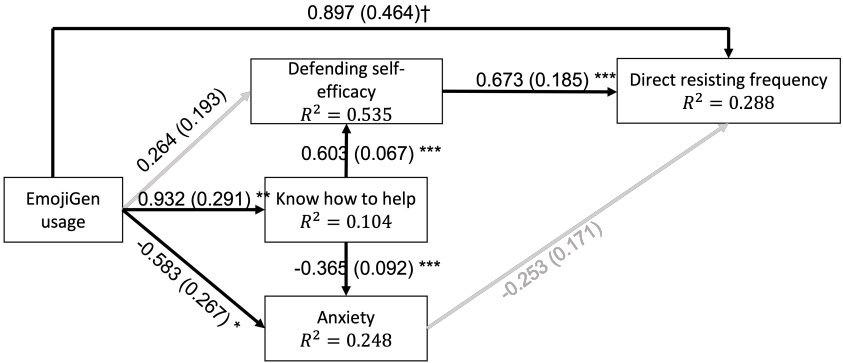

Fig. 6. The figure shows a structural equation model (SEM) to understand how EmojiGen usage leads to improved direct resisting frequency of bystanders. Solid black arrows represent statistically significant paths ($p < 0.05$), while gray arrows indicate non-significant paths. Path labels show the standardized regression coefficient, with the standard error in parentheses. Significance levels: $^{***}p < 0.001$, $^{**}p < 0.01$, $^{*}p < 0.05$, $^{\dagger}p < 0.10$ (marginal significance).

**Linear Regression.** We set the *direct resisting frequency* as the dependent variable, with *EmojiGen usage, defending self-efficacy, knowing how to help, anxiety,* and *personal responsibility* as predictors. The final regression model selected **defending self-efficacy** ($\beta = 0.392$, $p < 0.001$) and **EmojiGen usage** ($\beta = 0.230$, $p = 0.022$) as significant predictors. The model explained 26.6% of the variance in direct resistance frequency ($R^2 = 0.266$, $F(2, 85) = 15.38$, $p < 0.001$). Notably, other covariates such as anxiety, perceived responsibility, and knowledge of how to help were not retained in the final model.

**SEM Analysis.** Guided by the previous studies [7, 47, 51], which highlighted that a lack of ability, can increase anxiety and inhibit action [7, 47, 51], we considered *knowing how to help* as a foundational step that enables bystander intervention (*direct resisting frequency*) and reduces *anxiety*. Meanwhile, the sense of *knowing how to help* reflects participants' perception of their intervention skills, which is a decisive factor in *defending self-efficacy* [34]. Furthermore, *defending self-efficacy* has been identified as a key psychological driver of the direct intervention behaviors [7, 100]. In this analysis, personal responsibility was not a significant predictor of direct resisting frequency. Additionally, it did not change significantly due to EmojiGen usage, so it was not included in the SEM modeling. Fig. 6 shows the complete structure of the SEM analysis.

**Model fit.** The model demonstrated excellent fit to the data. The chi-square test was non-significant ($\chi^2(2) = 0.578$, $p = 0.749$), and all key fit indices exceeded recommended thresholds: $CFI = 1.000$, $TLI = 1.058$, $SRMR = 0.012$, $RMSEA = 0.000$ ($90\%CI[0.000, 0.145]$, $p = 0.793$). Additional fit indices such as $GFI = 1.000$ and McDonald's Fit Index ($MFI = 1.008$) further supported model adequacy.

**Key paths.** As shown in Table 5, several critical paths were significant. EmojiGen usage significantly increased bystanders' perceived knowledge of how to help ($\beta = 0.932$, $p = 0.001$), which subsequently predicted defending self-efficacy ($\beta = 0.603$, $p < 0.001$). Knowing how to help ($\beta = -0.365$, $p < 0.001$) and EmojiGen usage ($\beta = -0.583$, $p = 0.029$) both significantly reduced anxiety. As expected, defending self-efficacy positively predicted direct intervention ($\beta = 0.673$, $p < 0.001$). The direct effect of anxiety on resisting behavior was not statistically significant ($\beta = -0.253$, $p = 0.139$). The direct effect of EmojiGen usage on resisting behavior approached significance ($\beta = 0.897$, $p = 0.053$), suggesting a potential marginal direct influence.

Table 5. Regression coefficients with 1000 bootstrap samples.

| Outcome | Predictor | Estimate | Std. Error | z-value | p | 95% Confidence Interval Lower | Upper |
|---|---|---|---|---|---|---|---|
| Defending self-efficacy | **Know how to help** | 0.603 | 0.067 | 8.995 | < .001 | 0.467 | 0.736 |
| | EmojiGen usage | 0.264 | 0.193 | 1.364 | 0.172 | -0.115 | 0.662 |
| Know how to help | **EmojiGen usage** | 0.932 | 0.291 | 3.203 | 0.001 | 0.369 | 1.537 |
| Anxiety | **Know how to help** | -0.365 | 0.092 | -3.946 | < .001 | -0.555 | -0.145 |
| | **EmojiGen usage** | -0.583 | 0.267 | -2.187 | 0.029 | -1.119 | -0.055 |
| Direct resisting frequency | **Defending self-efficacy** | 0.673 | 0.185 | 3.635 | < .001 | 0.195 | 1.108 |
| | Anxiety | -0.253 | 0.171 | -1.478 | 0.139 | -0.710 | 0.095 |
| | **EmojiGen usage** | 0.897 | 0.464 | 1.933 | 0.053 | 0.291 | 1.842 |

*4.4.2  Mechanism Analysis for Direct Supportive Behaviors.* We set the frequency of direct support as the dependent variable, with *EmojiGen usage, defending self-efficacy, knowing how to help, anxiety,* and *personal responsibility* as candidate predictors. The final regression model selected **EmojiGen usage** ($\beta = 0.429$, $p < 0.001$) and **perceived responsibility** ($\beta = 0.192$, $p = 0.047$) as significant predictors. The model explained 24.9% of the variance in direct support frequency ($R^2 = 0.249$, $F(2, 85) = 14.07$, $p < 0.001$). Notably, other covariates such as defending self-efficacy, anxiety, and knowing how to help were not retained in the final model.

We did not construct an SEM for direct support for two key reasons. First, *personal responsibility*, the only psychological variable retained in the final regression model besides EmojiGen usage, was not significantly influenced by EmojiGen usage ($p = 0.131$), and thus cannot function as a mediator in the structural pathway from EmojiGen to support behavior. Second, unlike defending self-efficacy, personal responsibility is theorized to arise from psychological and contextual antecedents such as empathy, moral norms, or victim familiarity [10, 43, 60], which were not measured in the current study. As a result, there was no empirically supported or theoretically grounded latent structure that would justify building a path model. Therefore, we limited our analysis to regression, which directly reflects the predictive contribution of EmojiGen usage and perceived responsibility to support behavior.

*4.4.3  Summary.* Through a combination of regression and SEM, we identified the key psychological pathways by which EmojiGen facilitates bystander intervention in cyberbullying. For resisting behaviors, the SEM analysis revealed that EmojiGen enhances bystanders' sense of knowing how to help, which in turn boosts defending self-efficacy and reduces anxiety, ultimately leading to increased intervention. For supportive behaviors, only EmojiGen usage and personal responsibility were significant predictors, and no viable mediating structure was found. Together, these findings highlight that different mechanisms underlie resisting and supportive behaviors.

## 4.5  Qualitative Evidence

The qualitative data provides deeper insight into participants' experiences with EmojiGen, highlighting both its strengths and limitations.

*4.5.1  EmojiGen Lowered the Threshold for the Cyberbullying Intervention.* Some of the participants (N = 14) viewed it positively, perceiving it as an effective means to tackle cyberbullying in a constructive manner while reducing the effort and cognitive load involved in intervention. Specifically, EmojiGen can generate positive and inspiring comments to help bystanders create a positive online environment and support victims. User-37 noted, *"The AI was helpful in generating positive and uplifting comments based on the specific content of the post, which may have helped to combat the cyberbullying without actually needing to respond directly to the cyberbully."* In addition, EmojiGen assisted users in expressing their feelings without resorting to retaliatory insults, thereby avoiding the cycle of violence. User-7 shared, *"I think it is highly effective as it helps me to convey my feelings across, and not just insulting the cyberbully as that would be cyberbullying as well."*

Moreover, the AI tool significantly reduces the time and cognitive effort required for users to formulate responses, which makes them easier and more willing to comment. For example, User-34 mentioned, *"It generated statements for me to consider using and reduced my need to think of how to respond to the bully,"* while User-13 added, *"It makes it easier to comment, especially with cyberbullying. It reduces the time and cognitive load to think of what to say in the comments. I will comment more when I can use the AI."* Besides, User-65 stated that she was inspired by the content of EmojiGen to facilitate her own comment construction: *"I wanted to say something more in my own tone of voice, and using the AI-generated comment as inspiration".*

By generating responses based on users' selected emoticons, EmojiGen also alleviates the pressure of commenting. As User-4 stated, *"The AI is quite good at generating an appropriate response based on the emoji I selected. It can be used without my intervention, reducing my pressure in commenting."* Overall, AI-generated tools like EmojiGen not only help users quickly create suitable responses in cyberbullying situations but also ensure that these responses are encouraging and *"tactical" (User-57)*, effectively making *"the bar to stop cyberbullying is a lot lower compared to if I have to type the message myself to the bully (User-6)".*

### 4.5.2 Drawbacks Limiting the EmojiGen Potentially.

Conversely, several critical shortcomings were also reported by some participants (N = 18). They highlighted that AI-generated responses were often (1) generic and lacked the nuance necessary to address the complexities of cyberbullying or comfort victims effectively, (2) overly positive, and cannot effectively help them convey aggressive emotions to the bully.

More specifically, User-85 suggested that AI might inadvertently fuel cyberbullying with their generic responses by eliciting apathetic reactions from bullies, claiming *"I think it was interesting to utilize and form a framework, but I realized it does not capture some nuances in the original content. I do not tend to respond to generic comments when I choose to comment. I also do not tend to comment on cyberbullying as there is little avenue to follow up and truly stop the person IRL [in real life]. I think AI would actually fuel cyberbullying if used to generate responses (e.g. 'That's not cool man', 'That language is so 1800s.' would just solicit apathy from the bully)".* Similarly, User-10, User-38, User-48, and User-75 emphasized the ineffectiveness and generic nature of AI responses, which are often perceived as *"insincere and easily ignored by bullies"*. For instance, User-38 claimed: *"It seems too formal and plain when counteracting negative comments; it seems unoriginal and basic."* On the other hand, five users expressed their concern that the AI-generated content was inconsiderate and hard to comfort the cyberbullying victims, such as User-35 stated that *"The comments generated by AI are easy to use but they sound auto-generated and impersonal, whereas commenting by myself would have encouraged the person who made the post more, as they sound like a human put effort into thinking and typing something out."* User-62 and User-37 recalled the experience that they rephrased the generation content of EmojiGen because *" ai sounded really fake and its tone is very formal and not friend-like" (User-62)* and *"I wanted to add more content to it to fully express what I was feeling so that it would seem more genuine, heartfelt and thoughtful" (User-37).*

Furthermore, a participant evaluated AI-generated responses as overly positive, which required significant modification to be contextually appropriate. *"Although it did help me create relevant and moderately creative responses, I found the need to edit them because they initially felt overly positive" (User-30).* Three participants expressed a more aggressive attitude, believing that EmojiGen hindered them from attacking online abusers with aggressive language. In particular, User-1 stressed that *"some words cannot express all my mood. For example, I would like to say some bad words to these* *racists, but AI cannot do that.* This viewpoint was also supported by User-9 who mentioned that *"It can not express my anger."*

## 5 Discussion

This study demonstrates that EmojiGen can effectively increase bystanders' direct intervention behaviors in cyberbullying contexts by lowering the psychological and expressive barriers to action **(RQ1)**. Across both supportive and resistive interventions, EmojiGen increased intervention likelihood while improving perceived defending self-efficacy and knowing how to help, and reducing anxiety and workload **(RQ2)**. Beyond these behavioral effects, our findings reveal that different forms of direct intervention are enabled through distinct psychological pathways **(RQ3)**.

These findings enrich the CSCW community's understanding of cyberbullying intervention [31] (Section 5.1) and complement existing intervention methods (Section 5.2), including through prosocial design [8, 44, 54, 116], educational approaches [64, 95], and automatic methods [70, 113]. Our findings also suggest the design shortcomings and potential ethical issues of EmojiGen (Section 5.3). Ultimately, some design insights are provided for the development of future cyberbullying interventions (Section 5.4).

### 5.1 Re-examining the Bystander Intervention Model

The BIM posits that intervention requires a bystander to sequentially complete a decision-making chain: *noticing the event, interpreting it as an emergency, assuming personal responsibility, knowing how to help*, and *taking actions* [31]. Our findings help extend the understanding of this model. First, AI support can partially bypass or attenuate certain stages of the BIM. Participants using EmojiGen intervened more frequently despite no significant increase in personal responsibility, while their sense of knowing how to help was directly scaffolded by the system. This indicates that AI assistance can reduce reliance on internal motivational and skill-related prerequisites by providing immediate, actionable support. This finding invites a reconsideration of the rigidity and sequencing of BIM stages when intervention is technologically mediated.

Moreover, we found that supporting victims and resisting perpetrators rely on fundamentally different psychological mechanisms. Supportive actions were more closely associated with personal responsibility, whereas resistive actions depended more strongly on defending self-efficacy and intervention skills. One plausible explanation for this divergence is that personal responsibility often acts as a moral-driven motivator, prompting individuals to engage in prosocial behaviors such as comforting or encouraging victims [27, 55, 116]. These actions are generally perceived as having lower social risks and lower thresholds for action [37, 38]. In contrast, resisting perpetrators typically involves openly challenging them [38], a behavior that necessitates not only moral conviction but also sufficient confidence (defending self-efficacy) [24, 100], intervention skills [17, 30, 92, 110] and low anxiety [47]. These differences suggest that future research can design cyberbullying interventions based on different intervention objectives, rather than only based on BIM.

### 5.2 EmojiGen as an AI-Mediated Cyberbullying Intervention Complements Existing Approaches

EmojiGen could be deemed as a form of AI-mediated cyberbullying intervention (AI-MCI), where AI acts as a mediator [49], helping bystanders complete expressions and enabling victims or perpetrators to receive intervention signals. Although such designs are still rare in current cyberbullying interventions, this intervention method has some unique advantages that can compensate for existing intervention designs.

First, EmojiGen complements prior cyberbullying interventions driven by prosocial designs, such as empathy cultivation and responsibility enhancement [37, 116]. These methods focus on intrinsic driving factors, while EmojiGen directly empowers bystanders by providing them with capability scaffolding. As suggested by [17], relying solely on internal empathy or a sense of responsibility is difficult to compensate for their shortcomings in intervention skills. Therefore, EmojiGen lowers the threshold for bystanders to take action.

EmojiGen also supplements existing educational methods that focus on building intervention knowledge and defending self-efficacy through teaching and simulation [53, 64, 95, 119]. These methods often struggle to translate the learning experience into consistent direct action during real encounters [20, 87]. Although EmojiGen is not designed to increase intervention skills in the long term, it does provide a way to achieve flexible and just-in-time [90] intervention in real situations. This helps achieve the education-oriented approach's ultimate goal, enabling bystanders to address cyberbullying. However, this does not mean that EmojiGen will replace education-centered methods, because the goal of EmojiGen as a capability scaffold is to hope that bystanders will eventually gain intervention abilities [62] rather than always relying on AI.

Moreover, EmojiGen also complements the AI-driven automatic methods, such as automatic detection [70, 113], monitoring [4, 23], and enforcing social norms [22, 58, 108, 129]. EmojiGen does not rely entirely on trained AI models to perform automatic detection and management actions, as these methods can easily lead to false deletions or wrongful sanctions [113], making them difficult to adopt on a large scale. Instead, EmojiGen fully relies on the power of bystanders distributed in the online world, equipping these bystanders with the ability to stop cyberbullying, thereby forming a large-scale human intervention network. It more emphasizes the role of humans compared to AI automatic methods, preserving human judgment and agency [42], aligning with human-centered AI principles [46], and potentially preventing the risk of false deletions or wrongful sanctions [113].

Considering the advantages of EmojiGen and the nascent development of AI-MCI, we suggest that future research continue to explore the design paradigm of AI-MCI interventions, making fuller use of AI's capabilities to strengthen the role of bystanders and other network participants, and to continually mitigate the occurrence of cyberbullying.

## 5.3 Shortcomings and Ethical Considerations in EmojiGen's Design

Despite some achievements by EmojiGen, it still exhibits some shortcomings and potential ethical considerations that require ongoing attention and solutions from future research. First, EmojiGen did not significantly enhance bystanders' sense of personal responsibility, suggesting that lowering action barriers alone may be insufficient to strengthen intrinsic moral motivation. Responsibility is a deeply rooted construct shaped by empathy, moral norms, and social values [10, 37, 116], rather than by the perceived ease of action alone. This indicates that responsibility-oriented designs and AI-mediated skill support should be treated as complementary rather than interchangeable.

Second, participants perceived some AI-generated responses as generic or emotionally constrained. While constructive and non-escalatory language is important for effective intervention [102], overly formulaic responses may weaken perceived authenticity and emotional resonance [69]. This can reduce the comforting effect on victims or the deterrent impact on perpetrators, since when information is identified as AI-generated, it can lead to a diminished perception of the signal by the recipient [50, 69] and even raise ethical concerns. Specifically, victims often seek emotionally attuned and personalized signals of care when being targeted [50]. If AI-generated responses are perceived as generic or emotionally constrained, victims may interpret these interventions as lacking genuine empathy, potentially reinforcing feelings of alienation rather than alleviating harm. Additionally, when multiple bystanders rely on EmojiGen without sufficient personalization, it may lead to clusters of repetitive or stylistically uniform comments. In highly visible online settings, such convergence can undermine the perceived authenticity and social impact of support, making interventions appear performative rather than sincerely motivated. This suggests an important direction for future research that enables the intervention process to achieve bystanders' emotional satisfaction, while perpetrators and victims can receive appropriate signals.

## 5.4 Design Implications

*5.4.1 Adaptive Interfaces for Differentiated Intervention Pathways.* Supportive and resistive interventions are driven by distinct psychological mechanisms. Therefore, the system can be designed to automatically detect the status of the victim and perpetrator in the current post and provide corresponding nudging signals on the interface accordingly. Specifically, if the system infers that the victim particularly needs direct supportive assistance, the AI should generate a nudging signal on the interface to attract bystanders' attention and enhance personal responsibility or empathy. Meanwhile, if malicious information from the perpetrator is detected in the post, the system should highlight the location of this malicious information and generate pre-intervention content through AI to nudge bystanders to actively intervene.

*5.4.2 AI-Mediated Cyberbullying Intervention as an Integrative Support Framework.* Our findings suggest that AI-MCI can serve as a complementary approach, rather than a replacement, that reinforces existing prosocial and education-oriented intervention methods. Accordingly, future designs can integrate responsibility- and empathy-oriented nudges [8, 27, 116] with contextual AI support. For example, systems may first foreground victims' distress or highlight potential harm to activate moral engagement, and subsequently embed AI-MCI assistance to help bystanders articulate appropriate responses once they decide to act. Such integration enables bystanders to retain intrinsic motivation while receiving sufficient external support to overcome expressive and psychological barriers.

To promote broader adoption, AI-MCI systems can further embed educational content into real-world use. Intervention principles and exemplar responses developed through training or educational programs [64, 95] can be maintained as an evolving knowledge base, which AI systems dynamically retrieve and adapt during live incidents. This approach allows learning-oriented interventions to be operationalized in situ, transforming abstract knowledge into actionable support.

*5.4.3 Balancing Efficiency with Signal Authenticity.* While EmojiGen lowers the threshold for action, overly formulaic responses can weaken the perceived sincerity of the intervention [69]. One possible solution is to rely on the bystander's own data and use retrieval-augmented generation (RAG) [128] technology to establish a personalized intervention agent [80]. When EmojiGen generates intervention messages, it can first retrieve past comments from bystanders and learn the relevant content and emotion style [61, 86].

## 6 Limitation and Future Work

Although EmojiGen achieves the goal of promoting bystanders' direct intervention behaviors, we acknowledge several limitations in our study. The participant's cultural context, the nature of the post, and the specific type of cyberbullying can all influence the research outcomes. In our experiment, we controlled the subjective severity of posts and comments to evaluate our design, which might limit the generalizability. Future work should obtain broader insights from a larger sample and a wider range of post and comment types.

Second, the platform remained an experimentally controlled environment, and participants did not experience actual long-term social consequences, although SnapShare referred to prior research [37] that closely resembled existing social networking platforms and participants did not express doubts regarding the authenticity of the platform or the posts (see Section 3.5) under appropriate experiment guidance (Section 3.8). Additionally, we employed self-chosen nicknames, which reduced perceived social risk by weakening the linkage between intervention behavior and participants' real-world identities. In contrast to real-world platforms where profiles, social ties, and posting histories are often persistent and publicly visible, such pseudonymous settings may lower concerns about reputation damage or retaliation [94, 125]. As a result, participants in our study may have perceived direct intervention, especially resistive actions, as less socially costly than they would in fully identifiable environments. Future work could explore deploying similar AI-assisted intervention mechanisms in collaboration with large-scale social media platforms such as Facebook or X, enabling the examination of bystander intervention behaviors in fully organic settings.

Third, the novelty effect [39], which may have influenced participants' engagement with the EmojiGen feature, is another widely existing concern for novel system design. We cannot deny that users will use EmojiGen out of novelty, but it is also undeniable that defending self-efficacy has been successfully improved, and the results of SEM further indicate that direct resisting is primarily driven by defending self-efficacy. Meanwhile, majority of bystanders in our study made further edits to the EmojiGen-generated content, indicating that they were not merely experimenting with the feature out of curiosity but were actively engaging in independent decision-making when posting their comments. This evidence partially alleviates concerns about the novel effect. Nevertheless, we still recommend that future studies be conducted over a longer period to further overcome the influence of the novelty effect.

Forth, although we employed SEM analysis to explore the mediating role of perceptual factors, it typically benefits from larger samples to ensure stable parameter estimation. With a sample size of N = 90, our mediation analysis should be interpreted with caution. While the use of bootstrapping partially mitigates this concern, the reported paths should be viewed as indicative rather than conclusive.

Moreover, to avoid potential misinterpretation of emojis [6], our study carefully selected the participants' age groups and selected emojis. However, this may limit the applicability of our findings to a broader set of emojis, and the age restrictions may also impact the generalizability of the conclusions. Therefore, we recommend that future research explore a wider range of emojis and include participants from diverse age groups to derive more universally applicable conclusions.

Finally, the primary goal of promoting bystander intervention is to effectively stop further harm from bullies and provide comfort to victims through the active roles of bystanders. While EmojiGen's design draws on relevant findings from previous studies and encourages bystanders to engage in direct intervention, there is currently insufficient empirical evidence to demonstrate whether and how this direct support influences the actual emotions of bullies and victims. Therefore, we encourage future research to focus on providing empirical evidence from the perspectives of victims or bullies to demonstrate the effectiveness of AI in promoting cyberbullying intervention.

## 7  Conclusion

We introduced EmojiGen, an AI intervention tool that uses emoji-driven input to generate context-aware responses, effectively lowering the barrier to direct bystander intervention in cyberbullying. Our experimental results demonstrate that EmojiGen significantly increases both supportive and resistive interventions, enhances bystanders' defending self-efficacy and sense of knowing how to help, and reduces perceived workload and anxiety. However, it does not significantly influence responsibility and qualitative feedback highlights challenges in authenticity and emotional expressiveness. These findings underscore the value of AI in scaffolding real-time intervention skills, while also pointing to the need for future designs that integrate empathy-building mechanisms, support differentiated intervention strategies, and prioritize ethical and personalized communication. EmojiGen represents a meaningful step toward human-centered, AI-supported systems for fostering safer online communities.

## References

[1] Riyadh Tariq Kadhim Al-Ameedi, Mohanned Jassim Dakhils Al-Ghizzy, et al. 2022. Investigating cyberbullying in electronic communication: A descriptive study. *International Journal of English Language Studies* 4, 4 (2022), 97–106.

[2] Hassan Ali, Philipp Allgeuer, and Stefan Wermter. 2024. Comparing Apples to Oranges: LLM-powered Multimodal Intention Prediction in an Object Categorization Task. *arXiv preprint arXiv:2404.08424* (2024).

[3] Kimberley R Allison and Kay Bussey. 2016. Cyber-bystanding in context: A review of the literature on witnesses' responses to cyberbullying. *Children and Youth Services Review* 65 (2016), 183–194.

[4] Ashwaq Alsoubai, Xavier V Caddle, Ryan Doherty, Alexandra Taylor Koehler, Estefania Sanchez, Munmun De Choudhury, and Pamela J Wisniewski. 2022. Mosafely, is that sus? a youth-centric online risk assessment dashboard. In *Companion Publication of the 2022 Conference on Computer Supported Cooperative Work and Social Computing.* 197–200.

[5] Jenn Anderson, Mary Bresnahan, and Catherine Musatics. 2014. Combating weight-based cyberbullying on Facebook with the dissenter effect. *Cyberpsychology, Behavior, and Social Networking* 17, 5 (2014), 281–286.

[6] Inamul Azad, Sugandha Chhibber, and Azra Tajhizi. 2023. How Do Different Generations Communicate on Social Media? A Comparative Analysis of Language Styles, Emoji Usage, and Visual Elements. *Language, Technology, and Social Media* 1, 2 (2023), 86–97.

[7] Albert Bandura. 1977. Self-efficacy: toward a unifying theory of behavioral change. *Psychological review* 84, 2 (1977), 191.

[8] Kirstin Barchia and Kay Bussey. 2011. Predictors of student defenders of peer aggression victims: Empathy and social cognitive factors. *International Journal of Behavioral Development* 35, 4 (2011), 289–297.

[9] Julia Barlińska, Anna Szuster, and Mikołaj Winiewski. 2018. Cyberbullying among adolescent bystanders: Role of affective versus cognitive empathy in increasing prosocial cyberbystander behavior. *Frontiers in psychology* 9 (2018), 799.

[10] Sara Bastiaensens, Heidi Vandebosch, Karolien Poels, Katrien Van Cleemput, Ann DeSmet, and Ilse De Bourdeaudhuij. 2014. Cyberbullying on social network sites. An experimental study into bystanders' behavioural intentions to help the victim or reinforce the bully. *Computers in Human Behavior* 31 (2014), 259–271.

[11] Sara Bastiaensens, Heidi Vandebosch, Karolien Poels, Katrien Van Cleemput, Ann DeSmet, and Ilse De Bourdeaudhuij. 2015. 'Can I afford to help?'How affordances of communication modalities guide bystanders' helping intentions towards harassment on social network sites. *Behaviour & Information Technology* 34, 4 (2015), 425–435.

[12] Lindsay Blackwell, Jill Dimond, Sarita Schoenebeck, and Cliff Lampe. 2017. Classification and its consequences for online harassment: Design insights from heartmob. *Proceedings of the ACM on human-computer interaction* 1, CSCW (2017), 1–19.

[13] Boomerang. 2018. Respondable: Write Better Email. https://www.boomeranggmail.com/respondable/

[14] Sara Mota Borges Bottino, Cássio Bottino, Caroline Gomez Regina, Aline Villa Lobo Correia, and Wagner Silva Ribeiro. 2015. Cyberbullying and adolescent mental health: systematic review. *Cadernos de saude publica* 31 (2015), 463–475.

[15] Virginia Braun and Victoria Clarke. 2006. Using thematic analysis in psychology. *Qualitative research in psychology* 3, 2 (2006), 77–101.

[16] Nicholas Brody and Anita L Vangelisti. 2017. Cyberbullying: Topics, strategies, and sex differences. *Computers in Human Behavior* 75 (2017), 739–748.

[17] Shawn Meghan Burn. 2009. A situational model of sexual assault prevention through bystander intervention. *Sex roles* 60 (2009), 779–792.

[18] Angela Busacca and Melchiorre Alberto Monaca. 2023. Deepfake: Creation, Purpose, Risks. In *Innovations and Economic and Social Changes due to Artificial Intelligence: The State of the Art.* Springer, 55–68.

[19] Margaret Byrne, Rayner Kay Jin Tan, Dan Wu, Gifty Marley, Takhona Grace Hlatshwako, Yusha Tao, Jennifer Bissram, Sophie Nachman, Weiming Tang, Rohit Ramaswamy, et al. 2023. Prosocial interventions and health outcomes: a systematic review and meta-analysis. *JAMA network open* 6, 12 (2023), e2346789–e2346789.

[20] Elisa Cantone, Anna P Piras, Marcello Vellante, Antonello Preti, Sigrun Daníelsdóttir, Ernesto D'Aloja, Sigita Lesinskiene, Mathhias C Angermeyer, Mauro G Carta, and Dinesh Bhugra. 2015. Interventions on bullying and cyberbullying in schools: A systematic review. *Clinical practice and epidemiology in mental health: CP & EMH* 11, Suppl 1 M4 (2015), 58.

[21] David Carneros-Prado, Laura Villa, Esperanza Johnson, Cosmin C Dobrescu, Alfonso Barragán, and Beatriz García-Martínez. 2023. Comparative study of large language models as emotion and sentiment analysis systems: A case-specific analysis of GPT vs. IBM Watson. In *International Conference on Ubiquitous Computing and Ambient Intelligence*. Springer, 229–239.

[22] Eshwar Chandrasekharan, Chaitrali Gandhi, Matthew Wortley Mustelier, and Eric Gilbert. 2019. Crossmod: A cross-community learning-based system to assist reddit moderators. *Proceedings of the ACM on human-computer interaction* 3, CSCW (2019), 1–30.

[23] Jonathan P Chang, Charlotte Schluger, and Cristian Danescu-Niculescu-Mizil. 2022. Thread with caution: Proactively helping users assess and deescalate tension in their online discussions. *Proceedings of the ACM on human-computer interaction* 6, CSCW2 (2022), 1–37.

[24] Hong Chen, Yong Fang, Ling Wang, Yanjun Chen, and Cuiying Fan. 2025. The relationship between defending self-efficacy and defending behavior in cyberbullying: a moderated mediation model. *BMC psychology* 13, 1 (2025), 1–11.

[25] Justin Cheng, Michael Bernstein, Cristian Danescu-Niculescu-Mizil, and Jure Leskovec. 2017. Anyone can become a troll: Causes of trolling behavior in online discussions. In *Proceedings of the 2017 ACM conference on computer supported cooperative work and social computing*. 1217–1230.

[26] Charles Chiang and Diego Gomez-Zara. 2024. The Evolution of Emojis for Sharing Emotions: A Systematic Review of the HCI Literature. *arXiv preprint arXiv:2409.17322* (2024).

[27] Xiaowei Chu, Yujing Zhao, Xin Li, Sumin Yang, and Yuju Lei. 2024. The sense of responsibility and bystanders' prosocial behavior in cyberbullying: The mediating role of compassion and the moderating roles of moral outrage and moral disgust. *Cyberpsychology: Journal of Psychosocial Research on Cyberspace* 18, 3 (2024).

[28] Madeleine Clark and Kay Bussey. 2020. The role of self-efficacy in defending cyberbullying victims. *Computers in Human Behavior* 109 (2020), 106340.

[29] Henriette Cramer, Paloma De Juan, and Joel Tetreault. 2016. Sender-intended functions of emojis in US messaging. In *Proceedings of the 18th international conference on human-computer interaction with mobile devices and services*. 504–509.

[30] Robert Ervin Cramer, M Rosalie Mcmaster, Patricia A Bartell, and Marguerite Dragna. 1988. Subject competence and minimization of the bystander effect. *Journal of Applied Social Psychology* 18, 13 (1988), 1133–1148.

[31] John M Darley and Bibb Latané. 1968. Bystander intervention in emergencies: diffusion of responsibility. *Journal of personality and social psychology* 8, 4p1 (1968), 377.

[32] Pooja Datta, Dewey Cornell, and Francis Huang. 2016. Aggressive attitudes and prevalence of bullying bystander behavior in middle school. *Psychology in the Schools* 53, 8 (2016), 804–816.

[33] Anna Davidovic, Catherine Talbot, Catherine Hamilton-Giachritsis, and Adam Joinson. 2023. To intervene or not to intervene: young adults' views on when and how to intervene in online harassment. *Journal of Computer-Mediated Communication* 28, 5 (2023), zmad027.

[34] Edward L Deci and Richard M Ryan. 2013. *Intrinsic motivation and self-determination in human behavior*. Springer Science & Business Media.

[35] Simon Denny, Elizabeth R Peterson, Jaimee Stuart, Jennifer Utter, Pat Bullen, Theresa Fleming, Shanthi Ameratunga, Terryann Clark, and Taciano Milfont. 2015. Bystander intervention, bullying, and victimization: A multilevel analysis of New Zealand high schools. *Journal of school violence* 14, 3 (2015), 245–272.

[36] Ann DeSmet, Sara Bastiaensens, Katrien Van Cleemput, Karolien Poels, Heidi Vandebosch, and Ilse De Bourdeaudhuij. 2012. Mobilizing bystanders of cyberbullying: An exploratory study into behavioural determinants of defending the victim. *Annual Review of Cybertherapy and Telemedicine 2012* (2012), 58–63.

[37] Dominic DiFranzo, Samuel Hardman Taylor, Franccesca Kazerooni, Olivia D Wherry, and Natalya N Bazarova. 2018. Upstanding by design: Bystander intervention in cyberbullying. In *Proceedings of the 2018 CHI conference on human factors in computing systems*. 1–12.

[38] Kelly P Dillon and Brad J Bushman. 2015. Unresponsive or un-noticed?: Cyberbystander intervention in an experimental cyberbullying context. *Computers in Human Behavior* 45 (2015), 144–150.

[39] Dirk M Elston. 2021. The novelty effect. *Journal of the American Academy of Dermatology* 85, 3 (2021), 565–566.

[40] Sara Erreygers, Sara Pabian, Heidi Vandebosch, and Elfi Baillien. 2016. Helping behavior among adolescent bystanders of cyberbullying: The role of impulsivity. *Learning and Individual Differences* 48 (2016), 61–67.

[41] Paul Evans, Maarten Vansteenkiste, Philip Parker, Andrew Kingsford-Smith, and Sijing Zhou. 2024. Cognitive load theory and its relationships with motivation: a self-determination theory perspective. *Educational Psychology Review* 36, 1 (2024), 7.

[42] Lior Fink, Leorre Newman, and Uriel Haran. 2024. Let me decide: Increasing user autonomy increases recommendation acceptance. *Computers in Human Behavior* 156 (2024), 108244.

[43] Camilla Forsberg, Robert Thornberg, and Marcus Samuelsson. 2014. Bystanders to bullying: Fourth-to seventh-grade students' perspectives on their reactions. *Research Papers in Education* 29, 5 (2014), 557–576.

[44] Gianluca Gini. 2006. Social cognition and moral cognition in bullying: What's wrong? *Aggressive Behavior: Official Journal of the International Society for Research on Aggression* 32, 6 (2006), 528–539.

[45] Rebecca Godard and Susan Holtzman. 2022. The multidimensional lexicon of emojis: A new tool to assess the emotional content of emojis. *Frontiers in Psychology* 13 (2022), 921388.

[46] John D Gould and Clayton Lewis. 1985. Designing for usability: key principles and what designers think. *Commun. ACM* 28, 3 (1985), 300–311.

[47] JA Gray and N McNaughton. 2000. The psychology of Anxiety and Enquiry in to the functions of the septo hippocampus system.

[48] Hussam Habib, Maaz Bin Musa, Fareed Zaffar, and Rishab Nithyanand. 2019. To act or react: Investigating proactive strategies for online community moderation. *arXiv preprint arXiv:1906.11932* (2019).

[49] Jeffrey T. Hancock, Mor Naaman, and Karen Levy. 2020. AI-Mediated Communication: Definition, Research Agenda, and Ethical Considerations. *Journal of Computer-Mediated Communication* 25, 1 (2020), 89–100. https://doi.org/10.1093/jcmc/zmz022

[50] Hadassah Harland, Richard Dazeley, Hashini Senaratne, Peter Vamplew, Francisco Cruz, and Bahareh Nakisa. 2024. AI apology: A critical review of apology in AI systems. *arXiv preprint arXiv:2412.15787* (2024).

[51] Catherine A Hartley and Elizabeth A Phelps. 2012. Anxiety and decision-making. *Biological psychiatry* 72, 2 (2012), 113–118.

[52] Michael A Hedderich, Natalie N Bazarova, Wenting Zou, Ryun Shim, Xinda Ma, and Qian Yang. 2024. A Piece of Theatre: Investigating How Teachers Design LLM Chatbots to Assist Adolescent Cyberbullying Education. *arXiv preprint arXiv:2402.17456* (2024).

[53] Rose Hennessy Garza, Young Cho, Heather Hlavka, Lance Weinhardt, Tajammal Yasin, Sara Smith, Katharine Adler, Kacie Otto, and Paul Florsheim. 2023. A multi-topic bystander intervention program for upper-level undergraduate students: outcomes in sexual violence, racism, and high-risk alcohol situations. *Journal of interpersonal violence* 38, 15-16 (2023), 9395–9422.

[54] Frederick Herzberg, Bernard Mausner, and Barbara Bloch Snyderman. 2011. *The motivation to work*. Vol. 1. Transaction publishers.

[55] Martin L Hoffman. 1982. Development of prosocial motivation: Empathy and guilt. *The development of prosocial behavior* (1982), 281–313.

[56] Brett Holfeld. 2014. Perceptions and attributions of bystanders to cyber bullying. *Computers in Human Behavior* 38 (2014), 1–7.

[57] Thomas Holtgraves and Caleb Robinson. 2020. Emoji can facilitate recognition of conveyed indirect meaning. *PloS one* 15, 4 (2020), e0232361.

[58] Manoel Horta Ribeiro, Robert West, Ryan Lewis, and Sanjay Kairam. 2025. Post guidance for online communities. *Proceedings of the ACM on Human-Computer Interaction* 9, 2 (2025), 1–26.

[59] Liangjiecheng Huang, Weiqiang Li, Zikai Xu, Hongli Sun, Danfeng Ai, Yinfeng Hu, Shiqi Wang, Yu Li, and Yanyan Zhou. 2023. The severity of cyberbullying affects bystander intervention among college students: the roles of feelings of responsibility and empathy. *Psychology Research and Behavior Management* (2023), 893–903.

[60] Yun-yin Huang and Chien Chou. 2010. An analysis of multiple factors of cyberbullying among junior high school students in Taiwan. *Computers in human behavior* 26, 6 (2010), 1581–1590.

[61] Angel Hsing-Chi Hwang, Q. Vera Liao, Su Lin Blodgett, Alexandra Olteanu, and Adam Trischler. 2025. 'It was 80% me, 20% AI': Seeking Authenticity in Co-Writing with Large Language Models. *Proc. ACM Hum.-Comput. Interact.* 9, 2, Article CSCW122 (May 2025), 41 pages. https://doi.org/10.1145/3711020

[62] Shari L Jackson, Joseph Krajcik, and Elliot Soloway. 1998. The design of guided learner-adaptable scaffolding in interactive learning environments. In *Proceedings of the SIGCHI conference on Human factors in computing systems*. 187–194.

[63] Yanru Jia, Yuntena Wu, Tonglin Jin, and Lu Zhang. 2022. How are bystanders involved in cyberbullying? A latent class analysis of the Cyberbystander and their characteristics in different intervention stages. *International journal of environmental research and public health* 19, 23 (2022), 16083.

[64] Lisa M Jones, Kimberly J Mitchell, and Cheryl L Beseler. 2024. The impact of youth digital citizenship education: Insights from a cluster randomized controlled trial outcome evaluation of the be internet awesome (BIA) curriculum. *Contemporary School Psychology* 28, 4 (2024), 509–523.

[65] Anjuli Kannan, Karol Kurach, Sujith Ravi, Tobias Kaufmann, Andrew Tomkins, Balint Miklos, Greg Corrado, Laszlo Lukacs, Marina Ganea, Peter Young, et al. 2016. Smart reply: Automated response suggestion for email. In *Proceedings of the 22nd ACM SIGKDD international conference on knowledge discovery and data mining*. 955–964.

[66] Linda K Kaye, Helen J Wall, and Stephanie A Malone. 2016. "Turn that frown upside-down": A contextual account of emoticon usage on different virtual platforms. *Computers in Human Behavior* 60 (2016), 463–467.

[67] Franccesca Kazerooni, Samuel Hardman Taylor, Natalya N Bazarova, and Janis Whitlock. 2018. Cyberbullying bystander intervention: The number of offenders and retweeting predict likelihood of helping a cyberbullying victim. *Journal of Computer-Mediated Communication* 23, 3 (2018), 146–162.

[68] Vivien Kemp and Anthony R Henderson. 2012. Challenges faced by mental health peer support workers: peer support from the peer supporter's point of view. *Psychiatric rehabilitation journal* 35, 4 (2012), 337.

[69] Pranav Khadpe, Kimi Wenzel, George Loewenstein, and Geoff Kaufman. 2025. Explaining the Reputational Risks of AI-Mediated Communication: Messages Labeled as AI-Assisted Are Viewed as Less Diagnostic of the Sender's Moral Character. *arXiv preprint arXiv:2509.09645* (2025).

[70] Seunghyun Kim, Afsaneh Razi, Gianluca Stringhini, Pamela J Wisniewski, and Munmun De Choudhury. 2021. A human-centered systematic literature review of cyberbullying detection algorithms. *Proceedings of the ACM on Human-Computer Interaction* 5, CSCW2 (2021), 1–34.

[71] Ellen M Kraft and Jinchang Wang. 2009. Effectiveness of cyber bullying prevention strategies: A study on students' perspectives. *International Journal of Cyber Criminology* 3, 2 (2009), 513.

[72] Min Lan. 2022. Exploring a Self-paced Online Course Design, Learning Engagement, and Effectiveness on Anti-cyberbullying Topic for Adolescents in Hong Kong. In *Digital Communication and Learning: Changes and Challenges*. Springer, 107–122.

[73] Bibb Latané and John M Darley. 1970. The unresponsive bystander: Why doesn't he help? *(No Title)* (1970).

[74] Jungup Lee, Hyekyung Choo, Yijing Zhang, Hoi Shan Cheung, Qiyang Zhang, and Rebecca P Ang. 2025. Cyberbullying Victimization and Mental Health Symptoms among Children and Adolescents: A Meta-Analysis of Longitudinal Studies. *Trauma, Violence, & Abuse* (2025), 15248380241313051.

[75] Jungup Lee, JongSerl Chun, Jinyung Kim, Jieun Lee, and Serim Lee. 2021. A social-ecological approach to understanding the relationship between cyberbullying victimization and suicidal ideation in South Korean adolescents: The moderating effect of school connectedness. *International Journal of Environmental Research and Public Health* 18, 20 (2021), 10623.

[76] Yen-Fen Lee, Gwo-Jen Hwang, and Pei-Ying Chen. 2022. Impacts of an AI-based cha bot on college students' after-class review, academic performance, self-efficacy, learning attitude, and motivation. *Educational technology research and development* 70, 5 (2022), 1843–1865.

[77] Amanda Lenhart, Mary Madden, Aaron Smith, Kristen Purcell, Kathryn Zickuhr, and Lee Rainie. 2011. Teens, Kindness and Cruelty on Social Network Sites: How American Teens Navigate the New World of" Digital Citizenship". *Pew Internet & American Life Project* (2011).

[78] Li Li and Yue Yang. 2018. Pragmatic functions of emoji in internet-based communication—a corpus-based study. *Asian-Pacific Journal of Second and Foreign Language Education* 3 (2018), 1–12.

[79] Qing Li. 2007. Bullying in the new playground: Research into cyberbullying and cyber victimisation. *Australasian Journal of Educational Technology* 23, 4 (2007).

[80] Xiaopeng Li, Pengyue Jia, Derong Xu, Yi Wen, Yingyi Zhang, Wenlin Zhang, Wanyu Wang, Yichao Wang, Zhaocheng Du, Xiangyang Li, et al. 2025. A survey of personalization: From rag to agent. *arXiv preprint arXiv:2504.10147* (2025).

[81] Gary Lupyan and Rick Dale. 2016. Why are there different languages? The role of adaptation in linguistic diversity. *Trends in cognitive sciences* 20, 9 (2016), 649–660.

[82] D Lynn Hawkins, Debra J Pepler, and Wendy M Craig. 2001. Naturalistic observations of peer interventions in bullying. *Social development* 10, 4 (2001), 512–527.

[83] Robert D Lytle, Tabrina M Bratton, and Heather K Hudson. 2021. Bystander apathy and intervention in the era of social media. In *The emerald international handbook of technology-facilitated violence and abuse*. Emerald Publishing Limited, 711–728.

[84] Syed Mahbub, Eric Pardede, and ASM Kayes. 2021. Detection of harassment type of cyberbullying: A dictionary of approach words and its impact. *Security and Communication Networks* 2021, 1 (2021), 5594175.

[85] Candelaria I Mahlke, Ute M Krämer, Thomas Becker, and Thomas Bock. 2014. Peer support in mental health services. *Current opinion in psychiatry* 27, 4 (2014), 276–281.

[86] Navonil Majumder, Pengfei Hong, Shanshan Peng, Jiankun Lu, Deepanway Ghosal, Alexander Gelbukh, Rada Mihalcea, and Soujanya Poria. 2020. MIME: MIMicking emotions for empathetic response generation. *arXiv preprint arXiv:2010.01454* (2020).

[87] Faye Mishna, Charlene Cook, Michael Saini, Meng-Jia Wu, and Robert MacFadden. 2011. Interventions to prevent and reduce cyber abuse of youth: A systematic review. *Research on Social Work Practice* 21, 1 (2011), 5–14.

[88] Kathryn L Modecki, Jeannie Minchin, Allen G Harbaugh, Nancy G Guerra, and Kevin C Runions. 2014. Bullying prevalence across contexts: A meta-analysis measuring cyber and traditional bullying. *Journal of Adolescent Health* 55, 5 (2014), 602–611.

[89] Alberto Muñoz-Ortiz, Carlos Gómez-Rodríguez, and David Vilares. 2023. Contrasting linguistic patterns in human and llm-generated text. *arXiv preprint arXiv:2308.09067* (2023).

[90] Inbal Nahum-Shani, Shawna N Smith, Bonnie J Spring, Linda M Collins, Katie Witkiewitz, Ambuj Tewari, and Susan A Murphy. 2018. Just-in-time adaptive interventions (JITAIs) in mobile health: key components and design principles for ongoing health behavior support. *Annals of Behavioral Medicine* (2018), 1–17.

[91] Eryn J Newman, Maryanne Garry, Daniel M Bernstein, Justin Kantner, and D Stephen Lindsay. 2012. Nonprobative photographs (or words) inflate truthiness. *Psychonomic Bulletin & Review* 19 (2012), 969–974.

[92] Amanda B Nickerson, Ariel M Aloe, Jennifer A Livingston, and Thomas Hugh Feeley. 2014. Measurement of the bystander intervention model for bullying and sexual harassment. *Journal of adolescence* 37, 4 (2014), 391–400.

[93] Magdalena Obermaier, Nayla Fawzi, and Thomas Koch. 2016. Bystanding or standing by? How the number of bystanders affects the intention to intervene in cyberbullying. *New media & society* 18, 8 (2016), 1491–1507.

[94] Xinyu Pan, Yubo Hou, and Qi Wang. 2023. Are we braver in cyberspace? Social media anonymity enhances moral courage. *Computers in Human Behavior* 148 (2023), 107880.

[95] Ichhya Pant, Bee-Ah Kang, Rajiv Rimal, et al. 2023. Improving bystander self-efficacy to prevent violence against women through interpersonal communication using mobile phone entertainment education: Randomized controlled trial. *JMIR formative research* 7, 1 (2023), e38688.

[96] Mijeong Park, Ok Yeon Cho, and Jeong Sil Choi. 2023. A Pilot Study to Examine the Effects of a Workplace Cyberbullying Cognitive Rehearsal Mobile Learning Program for Head Nurses: A Quasi-Experimental Study. In *Healthcare*, Vol. 11. MDPI, 2041.

[97] Jan Pfetsch. 2016. Who is who in cyberbullying? Conceptual and empirical perspectives on bystanders in cyberbullying. *A social-ecological approach to cyberbullying* (2016), 121–149.

[98] Karina Polanco-Levicán and Sonia Salvo-Garrido. 2021. Bystander roles in cyberbullying: A mini-review of who, how many, and why. *Frontiers in psychology* 12 (2021), 676787.

[99] Joshua R Polanin, Dorothy L Espelage, and Therese D Pigott. 2012. A meta-analysis of school-based bullying prevention programs' effects on bystander intervention behavior. *School psychology review* 41, 1 (2012), 47–65.

[100] Virpi Pöyhönen, Jaana Juvonen, and Christina Salmivalli. 2010. What does it take to stand up for the victim of bullying? The interplay between personal and social factors. *Merrill-Palmer Quarterly (1982-)* (2010), 143–163.

[101] Tiziana Pozzoli and Gianluca Gini. 2010. Active defending and passive bystanding behavior in bullying: The role of personal characteristics and perceived peer pressure. *Journal of abnormal child psychology* 38 (2010), 815–827.

[102] Jeroen Pronk, Tjeert Olthof, Frits A Goossens, and Lydia Krabbendam. 2019. Differences in adolescents' motivations for indirect, direct, and hybrid peer defending. *Social Development* 28, 2 (2019), 414–429.

[103] Peinuan Qin, Zicheng Zhu, Naomi Yamashita, Yitian Yang, Keita Suga, and Yi-Chieh Lee. 2025. AI-Based Speaking Assistant: Supporting Non-Native Speakers' Speaking in Real-Time Multilingual Communication. *arXiv preprint arXiv:2505.01678* (2025).

[104] David Rodrigues, Marília Prada, Rui Gaspar, Margarida V Garrido, and Diniz Lopes. 2018. Lisbon Emoji and Emoticon Database (LEED): Norms for emoji and emoticons in seven evaluative dimensions. *Behavior research methods* 50 (2018), 392–405.

[105] Anurag Sarkar, Shalabh Agarwal, Abir Ghosh, and Asoke Nath. 2015. Impacts of social networks: A comprehensive study on positive and negative effects on different age groups in a society. *International Journal* 3, 5 (2015), 177–190.

[106] Hannah L Schacter, Shayna Greenberg, and Jaana Juvonen. 2016. Who's to blame?: The effects of victim disclosure on bystander reactions to cyberbullying. *Computers in human behavior* 57 (2016), 115–121.

[107] Charlotte Schluger, Jonathan P Chang, Cristian Danescu-Niculescu-Mizil, and Karen Levy. 2022. Proactive moderation of online discussions: Existing practices and the potential for algorithmic support. *Proceedings of the ACM on Human-Computer Interaction* 6, CSCW2 (2022), 1–27.

[108] Joseph Seering, Manas Khadka, Nava Haghighi, Tanya Yang, Zachary Xi, and Michael Bernstein. 2024. Chillbot: Content moderation in the backchannel. *Proceedings of the ACM on Human-Computer Interaction* 8, CSCW2 (2024), 1–26.

[109] Ashish Sharma, Inna W Lin, Adam S Miner, David C Atkins, and Tim Althoff. 2023. Human–AI collaboration enables more empathic conversations in text-based peer-to-peer mental health support. *Nature Machine Intelligence* 5, 1 (2023), 46–57.

[110] R Lance Shotland and William D Heinold. 1985. Bystander response to arterial bleeding: helping skills, the decision-making process, and differentiating the helping response. *Journal of personality and social psychology* 49, 2 (1985), 347.

[111] Robert Slonje, Peter K Smith, and Ann Frisén. 2013. The nature of cyberbullying, and strategies for prevention. *Computers in human behavior* 29, 1 (2013), 26–32.

[112] Peter K Smith and James O'Higgins Norman. 2021. The Wiley Blackwell handbook of bullying: A comprehensive and international review of research and intervention. (2021).

[113] Devin Soni and Vivek K Singh. 2018. See no evil, hear no evil: Audio-visual-textual cyberbullying detection. *Proceedings of the ACM on Human-Computer Interaction* 2, CSCW (2018), 1–26.

[114] Charles D Spielberger, Fernando Gonzalez-Reigosa, Angel Martinez-Urrutia, Luiz FS Natalicio, and Diana S Natalicio. 1971. The state-trait anxiety inventory. *Revista Interamericana de Psicologia/Interamerican journal of psychology* 5, 3 & 4 (1971).

[115] John Sweller. 2011. Cognitive load theory. In *Psychology of learning and motivation*. Vol. 55. Elsevier, 37–76.

[116] Samuel Hardman Taylor, Dominic DiFranzo, Yoon Hyung Choi, Shruti Sannon, and Natalya N Bazarova. 2019. Accountability and empathy by design: Encouraging bystander intervention to cyberbullying on social media. *Proceedings of the ACM on Human-Computer Interaction* 3, CSCW (2019), 1–26.

[117] Robert Thornberg and Tomas Jungert. 2013. Bystander behavior in bullying situations: Basic moral sensitivity, moral disengagement and defender self-efficacy. *Journal of adolescence* 36, 3 (2013), 475–483.

[118] Robert Thornberg, Linda Wänström, Jun Sung Hong, and Dorothy L Espelage. 2017. Classroom relationship qualities and social-cognitive correlates of defending and passive bystanding in school bullying in Sweden: A multilevel analysis. *Journal of school psychology* 63 (2017), 49–62.

[119] Pamela Tozzo, Oriana Cuman, Eleonora Moratto, and Luciana Caenazzo. 2022. Family and educational strategies for cyberbullying prevention: A systematic review. *International Journal of Environmental Research and Public Health* 19, 16 (2022), 10452.

[120] Seda Gökçe Turan. 2021. Deepfake and digital citizenship: A long-term protection method for children and youth. In *Deep fakes, fake news, and misinformation in online teaching and learning technologies*. IGI Global, 124–142.

[121] Kathleen Van Royen, Karolien Poels, Heidi Vandebosch, and Philippe Adam. 2017. "Thinking before posting?" Reducing cyber harassment on social networking sites through a reflective message. *Computers in human behavior* 66 (2017), 345–352.

[122] Jessica Vitak, Kalyani Chadha, Linda Steiner, and Zahra Ashktorab. 2017. Identifying women's experiences with and strategies for mitigating negative effects of online harassment. In *Proceedings of the 2017 ACM Conference on Computer Supported Cooperative Work and Social Computing*. 1231–1245.

[123] Shaofeng Wang, Zhuo Sun, and Ying Chen. 2023. Effects of higher education institutes' artificial intelligence capability on students' self-efficacy, creativity and learning performance. *Education and Information Technologies* 28, 5 (2023), 4919–4939.

[124] Elizabeth Whittaker and Robin M Kowalski. 2015. Cyberbullying via social media. *Journal of school violence* 14, 1 (2015), 11–29.

[125] Leping You and Yu-Hao Lee. 2019. The bystander effect in cyberbullying on social network sites: Anonymity, group size, and intervention intentions. *Telematics and Informatics* 45 (2019), 101284.

[126] Yuval Zak, Yisrael Parmet, and Tal Oron-Gilad. 2020. Subjective Workload assessment technique (SWAT) in real time: Affordable methodology to continuously assess human operators' workload. In *2020 IEEE International Conference on Systems, Man, and Cybernetics (SMC)*. IEEE, 2687–2694.

[127] Hui Zhang. 2024. Cognitive load as a mediator in self-efficacy and English learning motivation among vocational college students. *PloS one* 19, 11 (2024), e0314088.

[128] Ruichen Zhang, Hongyang Du, Yinqiu Liu, Dusit Niyato, Jiawen Kang, Sumei Sun, Xuemin Shen, and H Vincent Poor. 2024. Interactive AI with retrieval-augmented generation for next generation networking. *IEEE Network* (2024).

[129] Lei Zheng, Christopher M Albano, Neev M Vora, Feng Mai, and Jeffrey V Nickerson. 2019. The roles bots play in Wikipedia. *Proceedings of the ACM on Human-Computer Interaction* 3, CSCW (2019), 1–20.

# A    Posts and Comments

This appendix presents the textual content of the social media posts and comments used as experimental stimuli in our study. To provide transparency while preserving ecological validity, we report the full text of each post and its associated comments on day 3, including usernames and timestamps as shown to participants during the experiment. All identifying information has been anonymized, and no images are included in this appendix.

The posts are organized into three categories: *appearance* (Table 4, which have already been displayed in the main text and will not be repeated here), *gender* (Table 6), and *race* (Table 7). Each category contains three posts, resulting in a total of nine posts. Within each post, the comment section includes one cyberbullying comment (i.e., flaming, harassment, or trolling), embedded among neutral or supportive comments to approximate realistic social media environments.

To ensure that each post could be combined with different cyberbullying types (as described in Section 3.7 and Table 3), we prepared three alternative cyberbullying comments for each post, corresponding to flaming, harassment, and trolling (see Table 8). These variants were presented to participants in different experimental arrangements, while the post content and surrounding neutral or supportive comments remained unchanged.

This appendix reports only the textual components of the stimuli. During the experiment, each post was presented together with a corresponding image to provide visual context; however, images are omitted here due to space constraints and ethical considerations.

Table 6. **Topic 2: Gender-related posts and comments.** This table includes posts involving gender identity, gender norms, and gendered expectations. Each post contains one gender-based cyberbullying comment embedded among neutral or supportive comments.

| User | Time | Content (Text-only version, omitting emojis and images) |
| --- | --- | --- |
| **Post G1 (Harassment)** | | |
| emily_love | 2024/7/29 02:06 | **[POST] I participated in a homosexuality themed parade today and realized that "love is a collision between two hearts, not two organs." This world needs more acceptance, and we should respect everyone's sexual orientation. #LoveIsLove #Acceptance #LGBTQ** |
| Lovestory1213 | 2024/7/30 15:29 | [Comment] This is inspiring! The world definitely needs more love and less hate |
| siddharth_kumar | 2024/7/30 13:06 | [Comment - Harassment] Homosexuality is so disgusting! Just imagining two men together is sickening |
| zhang_mei | 2024/7/30 01:07 | [Comment] This is such a beautiful message! Love is love |
| isha_patel | 2024/7/29 08:54 | [Comment] I'm so glad you shared this. Everyone deserves to be loved for who they are |
| **Post G2 (Trolling)** | | |
| qingheo702 | 2024/8/1 14:24 | **[POST] As a math major girl, my dream is to become a successful mathematician after graduation!!! #MathMajor #WomenInSTEM #DreamBig** |
| justin_88 | 2024/8/1 12:48 | [Comment - Trolling] The only math women are good at is addition and subtraction |
| Jack_Photography | 2024/8/1 04:44 | [Comment] You're such an inspiration! Don't let anyone stop you |
| miyu_yoshida | 2024/8/1 19:55 | [Comment] That's an amazing dream! Keep pushing forward, you can do it |
| sara_thequeen | 2024/8/1 14:45 | [Comment] Go for it! The world needs more brilliant female mathematicians |
| **Post G3 (Flaming)** | | |
| jack_bradley | 2024/8/8 12:44 | **[POST] Today I got a new manicure, it's so shiny and looks amazing! It's the best one I've had in the past six months! #Manicure #NailArt #ShinyNails** |
| wang_haoran | 2024/8/11 10:16 | [Comment] Great choice, they really suit you! |
| rahul_sharma | 2024/8/11 05:37 | [Comment] Super cool manicure, where did you get it done? |
| keeprunning111 | 2024/8/9 22:42 | [Comment - Flaming] Men with nail polish? Disgusting |

Table 7. **Topic 3: Race-related posts and comments.** This table presents posts involving race, ethnicity, and cultural background. Each post contains one race-related cyberbullying comment embedded among neutral or supportive comments to reflect realistic online interactions.

| User | Time | Content (Text-only version, omitting emojis and images) |
|---|---|---|
| **Post R1 (Harassment)** | | |
| haipeng_Beijing | 2024/7/18 20:59 | **[POST] From Asia to America! Heading to the US for my studies tomorrow. Wish me luck! #StudyAbroad #NewBeginnings #AsianStudent** |
| chen_yi | 2024/7/21 03:17 | [Comment] Go out there and show them what you're made of! So proud of you |
| rani_gupta | 2024/7/19 19:08 | [Comment] You're going to have an amazing experience. Good luck |
| johnny_quick | 2024/7/19 01:15 | [Comment] Safe travels! You'll do great in the US |
| Jackson_makeit | 2024/7/19 01:52 | [Comment - Harassment] Show me a photo with your tiny eyes |
| **Post R2 (Trolling)** | | |
| Minny_BeHappy | 2024/8/5 14:57 | **[POST] Never thought we could immigrate! We've successfully left Africa and landed in the US. Captured some beautiful clouds and sea from the plane! Hoping our new neighbors will like us, that my husband's job goes well, and that our kids make new friends at school! #ImmigrationJourney #NewLife #Hopeful** |
| rani_gupta | 2024/8/6 21:06 | [Comment] That's wonderful! I hope everything goes smoothly for you all |
| hannah_sun | 2024/8/6 08:54 | [Comment - Trolling] I hope your husband doesn't end up working with me, or I'll have to worry about getting punched every day haha |
| dave_black | 2024/8/6 02:49 | [Comment] Welcome to the US! Wishing you all the best in your new life |
| **Post R3 (Flaming)** | | |
| Parkash_gogogo | 2024/7/25 17:36 | **[POST] Looking to make some new friends! Anyone want to come over? I'll cook my best Indian food for us! #Friendship #HomeCooked #IndianCuisine** |
| alex_wang | 2024/7/27 19:01 | [Comment] Your cooking? Yes, please! What time should I be there? |
| sato_kazu | 2024/7/27 05:31 | [Comment - Flaming] Making friends is fine, but eating there? Indian food is the dirtiest, unclean food in the world |

Table 8. **Full mapping between posts and cyberbullying comment variants, totalling 9 posts and 27 cyberbullying comments.** For each post, three cyberbullying comment variants (harassment, trolling, and flaming) were prepared. In the experiment, each post was paired with one cyberbullying comment at a time, and different pairings were presented across experimental arrangements (see Table 3).

| Post ID | Post Content (Text-only version, omitting emojis and images) | Cyberbullying Type | Cyberbullying Comment |
|---|---|---|---|
| **Appearance** | | | |
| A1 | Just got this new spaghetti strap dress! It looks amazing, can't wait to wear it to school tomorrow. #NewOutfit #Fashion #SchoolStyle | Harassment | Wow, you're way too young to be dressing like that. Trying to seduce men or what? |
| | | Trolling | Wear that to school? It's too much~ You can wear even less. |
| | | Flaming | I bet you just want attention. So self-obsessed. |
| A2 | Is there any outfit that looks good on a girl who is 160cm and 80kg? #FashionStruggles #BodyPositivity #NeedHelp | Harassment | Maybe you should try to wear nothing. |
| | | Trolling | Wow, you're really asking for fashion tips? You should know that being this fat disqualifies you from talking about "beauty". |
| | | Flaming | You're fatter than the pig my grandma keeps. |
| A3 | Just had my first gym session! Hoping to get as strong as LeBron James in a year. #GymLife #FitnessGoals | Harassment | No girl would want to date someone as scrawny as you. |
| | | Trolling | Strong like LeBron? Maybe in your next life! |
| | | Flaming | Such a twig doesn't deserve to be called a man, more like a "tiny boy". |
| **Gender** | | | |
| G1 | I participated in a 'homosexuality' themed parade today and realized that "love is a collision between two hearts, not two organs." This world needs more acceptance, and we should respect everyone's sexual orientation.#LoveIsLove #Acceptance #LGBTQ | Harassment | Homosexuality is so disgusting! Just imagining two men together is sickening. |
| | | Trolling | Can't believe there are people who support such weirdness like homosexuality. |
| | | Flaming | Homosexuals should get off the Earth!!! |
| G2 | As a math major girl, my dream is to become a successful mathematician after graduation!!! #MathMajor #WomenInSTEM #DreamBig | Harassment | Now: I want to be a female mathematician. Ten years later: dancing at a strip club. |
| | | Trolling | The only math women are good at is addition and subtraction. |
| | | Flaming | Seriously? Women should just be housewives, not mathematicians. |
| G3 | Today I got a new manicure, it's so shiny and looks amazing! It's the best one I've had in the past six months. #Manicure #NailArt #ShinyNails | Harassment | Such a sissy move. Are you considering a gender change next? |
| | | Trolling | What's next, a dress and heels? Ladyboy. |
| | | Flaming | Men with nail polish? Disgusting. |
| **Race** | | | |
| R1 | From Asia to America! Heading to the US for my studies tomorrow. Wish me luck. #StudyAbroad #NewBeginnings #AsianStudent | Harassment | Show me a photo with your tiny eyes. |
| | | Trolling | Don't you Chinese people have any human rights? |
| | | Flaming | Why would anyone want to come here from China? You guys are so backwards. |
| R2 | Never thought we could immigrate! We've successfully left Africa and landed in the US. Captured some beautiful clouds and sea from the plane. Hoping our new neighbors will like us, that my husband's job goes well, and that our kids make new friends at school. #ImmigrationJourney #NewLife #Hopeful | Harassment | Looks like we have to watch out for one more nigger. |
| | | Trolling | I hope your husband doesn't end up working with me, or I'll have to worry about getting punched every day haha~. |
| | | Flaming | Welcome! Barbaric and stupid chimpanzees. |
| R3 | Looking to make some new friends! Anyone want to come over? I'll cook my best Indian food for us. #Friendship #HomeCooked #IndianCuisine | Harassment | Feel like I'm walking past a pigsty whenever I pass my neighbor's place. |
| | | Trolling | Indians really smell because they can go a month without a bath. |
| | | Flaming | Making friends is fine, but eating there? Indian food is the dirtiest, unclean food in the world. |